\documentclass[a4paper,11pt]{article}
\pdfoutput=1 

\usepackage{jcappub} 
\usepackage{cases}
\usepackage{ctable}
\usepackage{multirow}
\usepackage{amssymb} 

\title{Cluster abundance in chameleon $f(R)$ gravity I: toward an accurate halo mass function prediction}

\author[a]{Matteo~Cataneo,}
\author[a,b,c,d,e]{David~Rapetti,}
\author[f]{Lucas~Lombriser,}
\author[g]{Baojiu~Li}

\affiliation[a]{Dark Cosmology Centre, Niels Bohr Institute, University of Copenhagen, Juliane Maries Vej 30, 2100 Copenhagen, Denmark}
\affiliation[b]{Faculty of Physics, Ludwig-Maximilians-Universit\"at, Scheinerstr. 1, 81679 Munich, Germany}
\affiliation[c]{Excellence Cluster Universe, Boltzmannstr. 2, 85748 Garching, Germany}
\affiliation[d]{Center for Astrophysics and Space Astronomy, Department of Astrophysical and Planetary Science, University of Colorado, Boulder, C0 80309, USA}
\affiliation[e]{NASA Ames Research Center, Moffett Field, CA 94035, USA}
\affiliation[f]{Institute for Astronomy, University of Edinburgh, Royal Observatory, Blackford Hill, Edinburgh, EH9~3HJ, U.K.}
\affiliation[g]{Institute for Computational Cosmology, Department of Physics, Durham University, South Road, Durham DH1 3LE, UK }

\emailAdd{matteoc@dark-cosmology.dk}

\newcommand{\tsec}[1]{Sec.~\ref{#1}}

\newcommand{\bq}{\begin{equation}}
\newcommand{\eq}{\end{equation}}
\newcommand{\bqa}{\begin{eqnarray}}
\newcommand{\eqa}{\end{eqnarray}}

\newcommand{\rmd}{\ensuremath{\mathrm{d}}}

\newcommand{\Msunh}{M_\odot /h}
\newcommand{\hMpc}{h^{-1}~\text{Mpc}}

\newcommand{\Sm}{S_{\rm m}}
\newcommand{\rhom}{\rho_{\rm m}}
\newcommand{\rhomb}{\bar{\rho}_{\rm m}}
\newcommand{\drhom}{\delta\rhom}
\newcommand{\Om}{\Omega_{\rm m}}
\newcommand{\Olam}{\Omega_{\Lambda}}

\newcommand{\dfR}{\delta f_R}
\newcommand{\dR}{\delta R}

\newcommand{\RTH}{r_{\rm th}}
\newcommand{\rhoin}{\rho_{\rm in}}

\newcommand{\yhal}{y_{\rm h}}
\newcommand{\yenv}{y_{\rm env}}
\newcommand{\deltac}{\delta_{\rm c}}
\newcommand{\deltacLCDM}{\delta_{\rm c}^{\Lambda}}

\newcommand{\denv}{\delta_{\rm env}}
\newcommand{\dceff}{\delta_{\rm c}^{\rm eff}}
\newcommand{\denvpeak}{\delta_{\rm env}^{\rm peak}}

\definecolor{mygreen}{rgb}{0,0.6,0}
\definecolor{mymauve}{rgb}{0.58,0,0.82}

\abstract{
We refine the mass and environment dependent spherical collapse model of chameleon $f(R)$ gravity by calibrating a phenomenological correction inspired by the parameterized post-Friedmann framework against high-resolution $N$-body simulations. We employ our method to predict the corresponding modified halo mass function, and provide fitting formulas to calculate the enhancement of the $f(R)$ halo abundance with respect to that of General Relativity (GR) within a precision of $\lesssim 5\%$ from the results obtained in the simulations. Similar accuracy can be achieved for the full $f(R)$ mass function on the condition that the modeling of the reference GR abundance of halos is accurate at the percent level. We use our fits to forecast constraints on the additional scalar degree of freedom of the theory, finding that upper bounds competitive with current Solar System tests are within reach of cluster number count analyses from ongoing and upcoming surveys at much larger scales. Importantly, the flexibility of our method allows also for this to be applied to other scalar-tensor theories characterized by a mass and environment dependent spherical collapse.
}

\begin{document}
\maketitle
\flushbottom


\section{Introduction} \label{sec:intro}

The abundance of galaxy clusters depends on the growth rate of cosmic structures as well as on the expansion history of the universe. This makes it a powerful probe of cosmology as a function of redshift, and particularly suited to investigate the nature of dark energy and deviations from General Relativity (GR)~\cite{Albrecht:2006um,Rapetti:2009ri}. Current and upcoming galaxy cluster surveys, such as the Dark Energy Survey (DES)~\cite{Abbott:2005bi}, the extended Roentgen Survey with an Imaging Telescope Array (eROSITA)~\cite{Merloni:2012uf}, the South Pole Telescope Third-Generation survey (SPT-3G)~\cite{Benson:2014qhw}, the Large Synoptic Survey Telescope (LSST)~\cite{Abell:2009aa} and {\it Euclid}~\cite{Laureijs:2011gra}, will detect an unprecedented number of these objects covering two orders of magnitude in mass ($M \sim 10^{13.5} - 10^{15.5} \, \Msunh$) for redshifts $z \lesssim 2$, with accurate calibration of the mass-observable relations down to a few percent. In order to take full advantage of this wealth of data, numerical and theoretical predictions of the mass distribution of virialized structures (also known as halo mass function) must reach a similar level of precision. Extensive effort has gone into modeling and calibrating this fully nonlinear observable in the standard cosmological constant plus Cold Dark Matter ($\Lambda$CDM) paradigm (e.g.~\cite{Maggiore:2009rv,Maggiore:2009rw,Corasaniti:2011dr,Corasaniti:2010zt,Sheth:1999mn,Jenkins:2000bv,Tinker:2008ff,Tinker:2010my,Crocce:2009mg,Manera:2009ak,Warren:2005ey,Reed:2012ih,Lukic:2007fc,Watson:2012mt,Despali:2015yla,Bocquet:2015pva}), and some work in this direction has been carried out for alternative dark energy models and gravity theories (e.g.~\cite{Barreira:2013eea,Barreira:2013xea,Barreira:2014kra,Bhattacharya:2010wy,Cui:2012is,Brax:2012sy,Schmidt:2008tn,Li:2011uw,Lombriser:2013wta,Lombriser:2013eza,Kopp:2013lea,Achitouv:2015yha}).

In this paper, we focus on the class of scalar-tensor theories known as $f(R)$ gravity (for reviews see e.g.~\cite{Sotiriou:2008rp,DeFelice:2010aj}), where the standard Einstein-Hilbert action is replaced by a general nonlinear function of the Ricci scalar $R$. The function $f(R)$ can be adjusted to mimic the $\Lambda$CDM expansion history, which in turn limits deviations from the standard model only to the growth of structure on both linear and nonlinear scales due to the fifth force mediated by the new scalar degree of freedom, known as \textit{scalaron}~\cite{STAROBINSKY198099,Oyaizu:2008sr,Li:2011vk,Llinares:2013qbh,Puchwein:2013lza,Pogosian:2007sw}. Constraints from local experiments~\cite{Will:2005va} are only consistent with functional forms that display the so-called \textit{chameleon} screening mechanism~\cite{Khoury:2003rn}. This ensures that force modifications are suppressed and GR is recovered for structures with deep potential wells, as the Solar System and the Galaxy~\cite{Hu:2007nk,Brax:2008hh}. However, the same coupling between the scalaron and the standard matter fields responsible for the chameleon mechanism may lead to catastrophic particle production in the early universe prior to Big Bang Nucleosynthesis (BBN), which can only be alleviated through fine tuning of the scalaron initial conditions~\cite{Erickcek:2013oma,Erickcek:2013dea}. In addition, the scalaron amplitude has been strongly constrained on small scales and late times using unscreened local dwarf galaxies, with allowed values in the range $|f_{R0}| \lesssim 10^{-7}$ at 95.4\% confidence level~\cite{Jain:2012tn,Vikram:2013uba}. It is also worth noting that this relatively recent technique would still greatly benefit from further investigation on various relevant astrophysical systematic uncertainties. All in all, these results further support the observation that $f(R)$ theories are unlikely candidates for a fundamental theory of gravity. Nevertheless, they can still be regarded as effective theories at low redshifts and on cosmological scales, with measurable deviations from GR predictions of the large scale structure.

The first studies designed to test $f(R)$ gravity with cluster number counts constrained the allowed region of parameter space to $|f_{R0}| \lesssim 10^{-4}$ at 95.4\% confidence level~\cite{Schmidt:2009am,Lombriser:2010mp}. More recently, from the abundance of X-ray selected massive galaxy clusters and utilizing the conservative halo mass function (HMF) predictions of Schmidt et al.~(2009)~\cite{Schmidt:2008tn}, Cataneo et al.~(2015)~\cite{Cataneo:2014kaa} improved this upper bound by an order of magnitude. Upon accurate modeling of the nonlinear chameleon mechanism, and employing the same cluster abundance data, weak lensing mass calibration and cluster analysis~\cite{Mantz:2014paa} this constraint could be further reduced by about a factor of two. An even more interesting prospect comes from including lower mass objects ($M \sim 10^{13.5} \, \Msunh$) at low redshift ($z\sim0.1$) along with an improved mass calibration down to 5\%, which could further strengthen the upper limit to $|f_{R0}| \lesssim 10^{-6}$. Thus, cluster count constraints have the potential to be competitive with those set by astrophysical and local tests of gravity but on much larger scales~\cite{Lombriser:2011zw,Joyce:2014kja}.

To this end, we present a phenomenological modification of the spherical collapse model of Lombriser et al.~(2013)~\cite{Lombriser:2013wta}, which we calibrate against high-resolution $N$-body simulations to predict the relative abundance of halos in $f(R)$ gravity with respect to GR within a $5\%$ precision (see~\cite{Li:2011uw,Kopp:2013lea,Achitouv:2015yha} for alternative approaches; for recent applications of the theoretical mass function presented in~\cite{Kopp:2013lea,Achitouv:2015yha} see~\cite{Peirone:2016wca,Liu:2016xes}). This is the first in a series of two papers dedicated to accurately modeling, robustly analyzing and tightly constraining chameleon $f(R)$ gravity from the abundance of massive clusters. While here we develop an accurate model of the $f(R)$ mass function, observational constraints will be presented in the second paper of the series. In Sec.~\ref{sec:theory} we review the main aspects of $f(R)$ gravity including the chameleon screening. Sec.~\ref{sec:scfR} summarizes the spherical collapse approach of Lombriser et al.~(2013)~\cite{Lombriser:2013wta} and introduces our new parametrization to correct for residual inaccuracies in that model. The dark matter only cosmological simulations that we use to calibrate the new model are described in Sec.~\ref{sec:simulations}, and we present our halo mass function predictions in Sec.~\ref{sec:hmf}. We conclude in Sec.~\ref{sec:conclusions} with an outlook on possible extensions and applications of our results.


\section{Chameleon $f(R)$ gravity} \label{sec:theory}

The $f(R)$ gravity model is a simple extension of GR, in which the Einstein-Hilbert action in the Jordan frame is generalized to include an arbitrary nonlinear function of the scalar curvature $R$,
\bq
S = \frac{1}{2\kappa^2} \int \rmd^4 x \sqrt{-g} \left[ R + f(R) \right] + \Sm\left[\psi_{\rm m}; g_{\mu\nu}\right].
\label{eq:jordan}
\eq
Here $\kappa^2 \equiv 8\pi G$, $\Sm$ is the action of the ordinary matter fields $\psi_{\rm m}$, $g$ is the determinant of the metric tensor $g_{\mu\nu}$, and throughout $c=\hbar=1$. Obviously, GR with a cosmological constant is restored for $f = -2\Lambda$. In metric $f(R)$ gravity, the modified Einstein field equations can be derived by varying the action in Eq.~\eqref{eq:jordan} with respect to $g_{\mu\nu}$. In particular, in a matter dominated universe with a flat, spatially homogeneous and isotropic cosmological background, the Friedmann equation reads 	
\bq
H^2 + \frac{1}{6}f - \frac{\ddot{a}}{a}f_R + H\dot{f}_R = \frac{\kappa^2}{3}\rhomb,
\label{eq:Friedmann}
\eq
and the Ricci scalar
\bq
\bar{R} = 6\left( \frac{\ddot{a}}{a} + H^2 \right),
\label{eq:Ricci}
\eq
where overdots denote differentiation with respect to cosmic time, $a(t)$ is the scale factor, $H \equiv \dot{a}/a$ is the Hubble parameter and $\rhomb$ indicates the background density of matter. Overbars represent background quantities everywhere in the text. In Eq.~\eqref{eq:Friedmann} $f_R \equiv {\rm d} f / {\rm d} R$ is the new scalar degree of freedom of the theory, commonly known as scalaron. Following~\cite{Pogosian:2007sw}, Eqs.~\eqref{eq:Friedmann}-\eqref{eq:Ricci} can be combined to define the effective density 
\bq
\rho_{\rm eff} \equiv \frac{1}{\kappa^2}\left[ \frac{1}{2}(f_R R - f) - 3H^2 f_R - 3H\dot{f}_R \right],
\label{eq:effdensity}
\eq
which together with the continuity equation
\bq
\dot{\rho}_{\rm eff} + 3H\rho_{\rm eff}(1+w_{\rm eff}) = 0
\label{eq:continuity}
\eq
gives the equation of state for the effective fluid
\bq
w_{\rm eff} \equiv \frac{P_{\rm eff}}{\rho_{\rm eff}} = 
			-\frac{1}{3} - \frac{2}{3} \frac{H^2 f_R - H\dot{f}_R -\frac{1}{2}\ddot{f}_R - \frac{1}{6}f}
			{\frac{1}{6}f_R R - H^2f_R - H\dot{f}_R - \frac{1}{6}f}.
\label{eq:eos}
\eq
Although our screening refinement method presented in Sec.~\ref{sec:refinement} is applicable to any viable $f(R)$ or generalized chameleon model~\cite{Hu:2007nk,Pogosian:2007sw,Khoury:2003aq,Khoury:2003rn,Brax:2004qh,Brax:2011aw}, in the rest of this work we shall use the popular Hu-Sawicki functional form~\cite{Hu:2007nk}
\bq
f(R) = -2\Lambda \frac{R^n}{R^n + \mu^{2n}},
\label{eq:HSexact}
\eq
where $\Lambda > 0$, $\mu^2$ and $n > 0$ are free parameters. Upon defining $f_{R0} \equiv -2n\Lambda\mu^{2n}/\bar{R}_0^{n+1}$ and $\bar{R}_0 \equiv \bar{R}(z=0)$, in the high-curvature regime, $R \gg \mu^2$, Eq.~\eqref{eq:HSexact} simplifies to
\bq
f(R) = -2\kappa^2\bar{\rho}_\Lambda - \frac{f_{R0}}{n}\frac{\bar{R}_0^{n+1}}{R^n},
\label{eq:HSapprox}
\eq
with $\Lambda=\kappa^2\bar{\rho}_\Lambda$. For $|f_{R0}| \ll 1$ this approximation is valid at all redshifts $z \geqslant 0$ owing to the very different curvature values set by $\Lambda \sim \mathcal{O}(\bar{R}_0)$ and $\mu^2$. For this model, the authors in~\cite{Hu:2007nk} showed that Eq.~\eqref{eq:eos} presents $\mathcal{O}(|f_{R0}|)$ deviations from a cosmological constant. Considering that the abundance of galaxy clusters currently provides an upper bound of $|f_{R0}| \lesssim 10^{-5}$~\cite{Cataneo:2014kaa}, and that upcoming improvements could potentially bring this down to $|f_{R0}| \sim 10^{-6}$, we restrict our predictions to the range $10^{-6} \leqslant |f_{R0}| \leqslant 10^{-4}$. In this regime the background evolution closely mimics $\Lambda$CDM, and we can safely adopt $w_{\rm eff} = -1$.

The trace of the modified Einstein field equations gives the Klein-Gordon equation for the scalaron
\bq
\Box f_R = \frac{\partial V_{\rm eff}}{\partial f_R},
\label{eq:KleinGordon}
\eq
with the effective potential 
\bq
\frac{\partial V_{\rm eff}}{\partial f_R} = \frac{1}{3}\left( R - f_RR + 2f - \kappa^2\rhom \right).
\label{eq:EffPotential}
\eq
Interestingly, $V_{\rm eff}$ depends on the matter density $\rhom$, and for viable $f(R)$ models it presents a minimum at the GR expectation of $R=\kappa^2(\rhom + 4\bar{\rho}_\Lambda)$. Limiting our analysis to this class of models, for which $|f_R| \ll 1$ at all redshifts and $|f/R| \ll 1$ in the early universe~\cite{Hu:2007nk,Pogosian:2007sw}, in the quasi static approximation~\cite{Noller:2013wca,Hojjati:2012rf,Oyaizu:2008sr} Eq.~\eqref{eq:KleinGordon} reduces to the Poisson-type equation
\bq
\nabla^2 \dfR = \frac{a^2}{3}\left[ \dR(f_R) - \kappa^2 \drhom \right],
\label{eq:PoissonfR}
\eq
where coordinates are comoving and fluctuations are obtained removing the background, i.e. $\dfR = f_R(R) - f_R(\bar{R})$, $\dR = R - \bar{R}$, and $\drhom = \rhom - \rhomb$. We also define the potential $\Psi$ as the time-time metric perturbation $2\Psi \equiv \delta g_{00}/g_{00}$ in the longitudinal gauge. The evolution of $\Psi$ is coupled to the matter density and curvature fluctuations through the modified Poisson equation
\bq
\nabla^2 \Psi = \frac{2\kappa^2}{3}a^2\drhom - \frac{a^2}{6} \dR(f_R).
\label{eq:Poisson}
\eq
The system of Eqs.~\eqref{eq:PoissonfR}-\eqref{eq:Poisson} controls the growth of structure, with modifications with respect to GR sourced by how differently curvature responds to matter due to the nonlinear term $\dR(f_R)$. This effectively corresponds to an additional fifth force with a range given by the inverse mass of the scalaron, as we shall show in the next section. 


\subsection{Large- and small-field regimes}\label{sec:LargeSmall}

For viable $f(R)$ models, we can approximate the mass of the scalar field as
\bq
m_{f_R}^2 = \frac{\partial^2 V_{\rm eff}}{\partial f_R^2} \approx \frac{1}{3f_{RR}} \equiv \left( \frac{2\pi}{\lambda_{\rm C}} \right)^2,
\label{eq:fRmass}
\eq
where we also introduce the Compton wavelength $\lambda_{\rm C}$. The latter defines how far the field can propagate from the source. To gain valuable insight into the solutions to Eqs.~\eqref{eq:PoissonfR}-\eqref{eq:Poisson}, we use a spherically symmetric top-hat overdensity embedded in the cosmological background with constant radius $\RTH$ and mass $M=4\pi\RTH^3\drhom/3$. Following~\cite{Hu:2007nk}, we also define the effective mass
\bq
M_{\rm eff} \equiv 4\pi \int_0^{\RTH} (\drhom - \dR/\kappa^2)r^2 \rmd r, 
\label{eq:EffMass}
\eq
where $r$ denotes the physical distance from the center of the overdensity. By inspection of Eq.~\eqref{eq:PoissonfR}, $M_{\rm eff}$ can be interpreted as the mass sourcing the exterior scalar field fluctuations responsible for the fifth force. 

For a given overdensity, two limiting cases bracket the family of interior solutions for the scalar field: (i) the low-curvature solution, for $\dR \ll \kappa^2\drhom$; and (ii) the high-curvature solution, for $\dR \approx \kappa^2\drhom$. A necessary condition for (ii) is that the density must change on scales much longer than the local Compton wavelength implied by the high-curvature solution~\cite{Hu:2007nk}. For our top-hat profile, this condition is always violated at the boundary with the cosmological background, consequently part of the exterior must be at low curvature. In addition, Birkhoff's theorem no longer applies (see e.g.~\cite{Martino:2008ae}) and the exterior low-curvature solution can enter the overdensity, even if the condition above is satisfied. The thickness of this region inside the overdensity depends upon the size of the overdensity itself, its density contrast and the amplitude of the cosmological scalar field. Therefore, the field does not always locally minimize the potential, rather it minimizes the total energy of the system which also includes the gradient kinetic energy associated with the field profile.
 
In terms of Eq.~\eqref{eq:EffMass}, if the entire overdensity is in the low-curvature regime, then $M_{\rm eff} \approx M$. The opposite is true if the high-curvature solution holds everywhere within the overdensity except close to the boundary, i.e. only a thin shell of mass $M_{\rm eff} \ll M$ contributes to the field gradients outside the overdensity. Applying Gauss's theorem to Eq.~\eqref{eq:PoissonfR} and using the definition of Eq.~\eqref{eq:EffMass}, we can write an implicit solution for the field fluctuations at $\RTH$~\cite{Hu:2007nk}
\bq
\dfR(\RTH) = \frac{2}{3}\frac{\kappa^2}{8\pi}\frac{M_{\rm eff}}{\RTH}. 
\label{eq:implicit}
\eq
Hence, the low-curvature solution provides the upper bound
\bq
\dfR(\RTH) \leqslant \frac{2}{3}|\Psi_{\rm N}|, 
\label{eq:implicit}
\eq
where $|\Psi_{\rm N}| = \kappa^2 M/8\pi \RTH$ defines the Newtonian potential at the surface of the sphere. This gives us a method to predict qualitatively the interior field profile for an isolated object and at a fixed background value. In fact, since $\dfR \lesssim |\bar{f}_R|$ we have that
\bqa
|\bar{f}_R| \gg |\Psi_{\rm N}| \, &\Longrightarrow& \, \dR \ll \kappa^2 \drhom,\\
|\bar{f}_R| \ll |\Psi_{\rm N}| \, &\Longrightarrow& \, \dR \approx \kappa^2 \drhom,
\label{eq:regimes}
\eqa
which we shall refer to as \textit{large-field regime} and \textit{small-field regime} respectively. The mechanism responsible for recovering the high-curvature solution in the small-field regime is called \textit{chameleon screening}~\cite{Khoury:2003rn}.

First, let us consider the case of a background scalar field $|\bar{f}_{R}| \gg |\Psi_{\rm N}| \sim 10^{-5}$, where $\Psi_{\rm N}$ now refers to the typical depth of the Newtonian potential for galaxy clusters, which are the objects that we are interested in here. In this scenario field fluctuations are relatively small and curvature fluctuations can be linearized as~\cite{Chiba:2006jp}
\bq
\dR \approx \frac{\partial R}{\partial f_R} \bigg |_{\bar{R}} \dfR = 3\bar{m}_{f_R}^2 \dfR.
\label{eq:dRlinear}
\eq
The combination of Eq.~\eqref{eq:PoissonfR} and Eq.~\eqref{eq:Poisson}, together with the approximation of Eq.~\eqref{eq:dRlinear}, gives the following solution for the potential in Fourier space
\bq
k^2\Psi({\bf k}) = -\frac{\kappa^2}{2}\left( 1 + \frac{1}{3} \frac{k^2}{k^2 + \bar{m}_{f_R}^2 a^2} \right) a^2 \drhom({\bf k}).
\label{eq:PotentialFourier}
\eq
On scales $k \gg \bar{m}_{f_R} a$ gravitational forces exhibit $1/3$ enhancements compared to GR. In this limit, the nature of the additional interaction becomes even clearer for a point-mass with density $\drhom(r) = M\delta_D(r)/2\pi r^2$, where $M$ is the mass, $\delta_D(r)$ denotes a Dirac delta function, and $r$ is expressed in physical coordinates. This system is equivalent to that of a top-hat overdensity of constant radius $\RTH$, for distances $r > \RTH$. For this particular case, Eq.~\eqref{eq:PotentialFourier} in real space takes the form
\bq
\Psi(r) = -\frac{\kappa^2}{8\pi} \frac{M}{r} - \frac{\kappa^2}{24\pi} \frac{M}{r} e^{-\bar{m}_{f_R}(r-\RTH)},
\label{eq:PotentialReal}
\eq
where the first term is the standard Newtonian potential and the second term represents a Yukawa-like potential with range defined by the background scalaron mass~\cite{Hu:2007nk}. Plugging Eq.~\eqref{eq:PotentialReal} into Eq.~\eqref{eq:Poisson}, and using Eq.~\eqref{eq:dRlinear} gives the exterior solution for the scalar field
\bq
\dfR(r) = \frac{\kappa^2}{12\pi} \frac{M}{r}e^{-\bar{m}_{f_R}(r-\RTH)} \quad {\rm for} \; r>\RTH.
\label{eq:exteriorfR}
\eq
The interior solution for the field is obtained from Eq.~\eqref{eq:PoissonfR} noticing that curvature fluctuations can be neglected inside the overdensity ($\dR \ll \kappa^2\drhom$). In addition, we require the interior and exterior solutions to match at $r=\RTH$, as well as ${\rm d}f_R/{\rm d}r = 0$ at $r=0$ to avoid divergences. With these boundary conditions the solution to Eq.~\eqref{eq:PoissonfR} is
\bq
\dfR(r) = \frac{\kappa^2}{8\pi}\frac{M}{\RTH}\left( 1-\frac{1}{3}\frac{r^2}{\RTH^2} \right) \quad {\rm for} \; r<\RTH.
\label{eq:interiorfR}
\eq
Both Eq.~\eqref{eq:exteriorfR} and Eq.~\eqref{eq:interiorfR} are the Jordan frame equivalent of Eqs.~(29) and~(30) in~\cite{Khoury:2003rn}.

In the small-field regime, $|\bar{f}_{R}| \ll |\Psi_{\rm N}| \sim 10^{-5}$, and the scalaron is close to the minimum of the effective potential everywhere inside the overdensity except for a negligible thin-shell at the boundary. This case is characterized by curvature perturbations approaching the GR limit $\dR = \kappa^2 \drhom$, implying small field gradients, $|\nabla^2 \dfR| \ll \kappa^2 \drhom$, that highly suppress force modifications. Hence, the interior solution for the scalar field will be
\bq
f_{R,{\rm in}} \approx f_R^{min} \equiv f_{R0} \left[ \frac{\bar{R}_0}{\kappa^2(\rhom + 4\bar{\rho}_\Lambda)} \right]^{n+1},
\label{eq:ScalaronHighCurvature}
\eq
which gives $|f_R| \ll |f_{R0}|$ for $\rhom \gg \rhomb$. Outside the overdensity the field moves towards the cosmological background with gradients negligible compared to the standard gravitational acceleration. In this regime, Eq.~\eqref{eq:Poisson} simply becomes the usual Poisson equation, and Eq.~\eqref{eq:PotentialReal} retains only the standard Newtonian contribution.

For $|\bar{f}_{R}| \sim |\Psi_{\rm N}| \sim 10^{-5}$, the exterior high-curvature solution can penetrate within the overdensity for a depth $\Delta r \lesssim \RTH$, effectively screening the interior and recovering GR at radii $r < \RTH - \Delta r$. In the next section, we shall estimate the thickness of this shell for our spherical top-hat overdensity with a method that includes the large- and small-field regimes as limiting cases, for $\Delta r \gtrsim \RTH$ and $\Delta r \ll \RTH$ respectively.  


\subsection{Intermediate regime}\label{sec:Intermediate}

In this section we follow the treatment presented in~\cite{Khoury:2003rn} for the estimation of the radial profile of a chameleon field $\phi$ in a compact object of radius $\RTH$ with constant matter density $\rho_\text{in}$ embedded in a background of homogenous density $\rho_\text{out}$. Using the conformal equivalence between $f(R)$ gravity and scalar-tensor theories (see e.g.~\cite{Pogosian:2007sw}), the authors in~\cite{Lombriser:2013wta} derive the thinkness of the shell required for the transition between the exterior and the interior fields both minimizing the effective potential of Eq.~\eqref{eq:EffPotential}. Denoting these two values $f_{R,\text{out}}$ and $f_{R,\text{in}}$ respectively, the extent of this region within the spherical top-hat overdensity is well approximated in the thin-shell regime $\Delta r/\RTH \ll 1$ by
\bq
\frac{\Delta r}{\RTH} \approx \frac{3}{\kappa^2 \rho_\text{in}} \frac{f_{R,\text{in}} - f_{R,\text{out}}}{\RTH^2},
\label{eq:shell}
\eq
where we also assumed $\RTH \ll \bar{\lambda}_\text{C}$. For a flat $\Lambda$CDM background, the interior and exterior values of the scalar fields minimizing $V_\text{eff}(f_R)$ are obtained directly from Eq.~\eqref{eq:ScalaronHighCurvature} as
\bq
f_{R,\text{in/out}} \approx f_{R0} \left[ \frac{1+4\frac{\Omega_\Lambda}{\Omega_\text{m}}}{\tilde{\rho}_\text{in/out}a^{-3}+4\frac{\Omega_\Lambda}{\Omega_\text{m}}} \right]^{n+1},
\label{eq:inout_fields}
\eq  
where $\tilde{\rho}_\text{in/out} \equiv \rho_\text{m,in/out}(a=1)/\rhomb(a=1)$, $\Om$ is the mean matter density today in units of the critical density $\bar{\rho}_{\rm cr}(a=1)$, and $\Olam=1-\Om$. Combining Eqs.~\eqref{eq:shell} and~\eqref{eq:inout_fields} we obtain the thickness of the thin-shell in terms of the background cosmology and the physical properties of the overdensity
\bq
\frac{\Delta r}{\RTH} \approx \frac{|f_{R0}|a^3}{\Om \tilde{\rho}_{\rm in}(H_0\RTH)^2}
					\left[ \left( \frac{1+4\frac{\Om}{\Olam}}{\tilde{\rho}_{\rm out}a^{-3} + 4\frac{\Om}{\Olam}} \right)^{n+1} 
					- \left( \frac{1+4\frac{\Om}{\Olam}}{\tilde{\rho}_{\rm in}a^{-3} + 4\frac{\Om}{\Olam}} \right)^{n+1}  \right],
\label{eq:shell_explicit}
\eq  
where $H_0$ denotes the present-day Hubble constant. Throughout, we will also use the equivalent dimensionless quantity $h=H_0/100~{\rm km/s/Mpc}$.

In the thin-shell limit, the approximate interior solution for the scalaron is

\begin{numcases}{f_R(r) \approx}
      f_{R,{\rm in}} & $r < r_0$, \nonumber \\
      f_{R,{\rm in}} - \frac{\kappa^2}{9}\rhoin \left( \frac{r^2}{2} + \frac{r_0^3}{r} - \frac{3}{2}r_0^2 \right) & $r_0 \leq r \leq \RTH$,
      \label{eq:interior_thinshell}
\end{numcases}
with $r_0 = \RTH - \Delta r$. Therefore, the magnitude of the additional fifth force $F$ for a unit mass at the surface of the overdensity is given by~\cite{Lombriser:2013wta,Khoury:2003rn}
\bq
F = \frac{1}{2} \nabla f_R \big |_{\RTH} \approx \frac{1}{3} F_N \left[ 3\frac{\Delta r}{\RTH} 
									- 3 \left( \frac{\Delta r}{\RTH} \right)^2 + \left( \frac{\Delta r}{\RTH} \right)^3 \right], 
\label{eq:fifth_force}
\eq
where $F_N = GM/\RTH^2$ is the Newtonian force. Although Eq.~\eqref{eq:fifth_force} is strictly valid only in the thin-shell limit, we extend it to include also the thick-shell regime, where $\Delta r/\RTH \gtrsim 1$ and $F=F_N/3$, by defining the ratio  
\bq
\mathcal{F} \equiv \frac{F}{F_N} = \frac{1}{3} \min \left( 3\frac{\Delta r}{\RTH} 
									- 3 \left( \frac{\Delta r}{\RTH} \right)^2 
									+ \left( \frac{\Delta r}{\RTH} \right)^3 , 1 \right),
\label{eq:force_ratio}
\eq
which provides an interpolation between the small-field regime $\mathcal{F} = 0$ and the large-field regime $\mathcal{F} = 1/3$.

Spherical collapse dynamics in $f(R)$ gravity is complicated by a breakdown of Birkhoff's theorem, inducing shell crossing where the low-curvature exterior solution enters the overdensity. In general, departures from GR lead to the dependence of structure formation on the environment, the halo substructure and the initial density profile~\cite{Martino:2008ae,Borisov:2011fu,Kopp:2013lea,Li:2011uw,Li:2011qda,Lombriser:2013wta}. Nevertheless, we adopt a simplified approach built on the assumption that the initial density profile also evolves as a spherical top-hat. In Secs.~\ref{sec:sctheory} and~\ref{sec:refinement} we will explain our method to fully account for nonlinear structure formation in $f(R)$ gravity within the spherical top-hat scenario, which also corrects for the inaccuracy of Eq.~\eqref{eq:force_ratio} in the thick-shell regime. 


\section{Spherical collapse in chameleon $f(R)$ gravity} \label{sec:scfR}

In Sec.~\ref{sec:sctheory} we first briefly summarize the spherical collapse model for $f(R)$ gravity presented in~\cite{Li:2011qda,Lam:2012fa,Lombriser:2013wta,Lombriser:2013eza}, and subsequently in Sec.~\ref{sec:refinement} we implement a novel correction into this formalism to account for the departures between the calculations in this simplified picture and those in fully non-linear cosmological $N$-body simulations.


\subsection{Mass and environment dependent spherical collapse}\label{sec:sctheory}


We adopt the spherical collapse model to describe halo formation in $f(R)$ gravity by approximating overdensities with spherically symmetric top hats that we evolve with the nonlinear continuity and Euler equations from an initial time to that of their collapse.
The chameleon screening effect can be incorporated in the spherical collapse calculation following Li \& Efstathiou~(2012)~\cite{Li:2011qda} (cf.~\cite{Borisov:2011fu}) by accounting for the mass and environment dependent gravitational force modification using the thin-shell thickness estimator of Khoury \& Weltman~(2004)~\cite{Khoury:2003rn} described in \tsec{sec:Intermediate}.
Further developments on the chameleon spherical collapse model and its applications to $f(R)$ gravity, the halo mass function, and the halo model have been developed in~\cite{Lam:2012fa,Lombriser:2013wta,Lombriser:2013eza}.
A review of these applications and a comparison to different approaches in modeling the nonlinear structure of chameleon models can be found in~\cite{Lombriser:2014dua}.

We define the physical radius of the top-hat overdensity as $\zeta(a)$. At an initial scale factor $a_{\rm i}\ll1$ this is given by $\zeta(a_{\rm i})=a_{\rm i}\RTH$, but it deviates from this simple linear relation when $a>a_{\rm i}$ due to its nonlinear evolution.
More specifically, the equation of motion of the spherical shell is given by~\cite{Schmidt:2008tn,Li:2011qda,Lombriser:2013wta}
\bq
\frac{\ddot{\zeta}}{\zeta} \simeq -\frac{\kappa^2}{6} \left( \rhomb - 2\bar{\rho}_{\Lambda} \right) - \frac{\kappa^2}{6} \left(1 + \mathcal{F} \right) \drhom \,,
\label{eq:shellmotion}
\eq
where the gravitational force modification $\mathcal{F}$ is given in Eq.~(\ref{eq:force_ratio}) and we replace $\Delta r / \RTH \rightarrow \Delta\zeta/\zeta$.
We define the dimensionless variable $y\equiv \zeta(a)/(a\RTH)$, and conservation of mass enclosed in the overdensity, $\rhomb a^3 \RTH^3 = \rhom \zeta^3$, yields $\tilde{\rho} = \rhom/\rhomb = y^{-3}$.
The evolution equation for $\yhal=\tilde{\rho}_{\rm in}^{-1/3}$ follows from Eq.~(\ref{eq:shellmotion}),
\bq
\yhal'' + \left[ 2 - \frac{3}{2} \Om(a) \right] \yhal' + \frac{1}{2} \Om(a) \left(1 + \mathcal{F} \right) \left( \yhal^{-3} - 1 \right) \yhal = 0 \,, \label{eq:yhal}
\eq
where the force enhancement is given by Eqs.~(\ref{eq:force_ratio}) with
\bq
\frac{\Delta \zeta}{\zeta} \approx \frac{|f_{R0}| a^{4+3n}}{\Om(H_0\RTH)^2} \yhal \left[ \left( \frac{1+4\frac{\Olam}{\Om}}{\yenv^{-3} + 4\frac{\Olam}{\Om} a^3} \right)^{n+1} - \left( \frac{1+4\frac{\Olam}{\Om}}{\yhal^{-3} + 4\frac{\Olam}{\Om} a^3} \right)^{n+1} \right] . \label{eq:thshsphcoll}
\eq
The environment $\yenv=\tilde{\rho}_{\rm out}^{-1/3}$ is assumed to evolve according to $\Lambda$CDM with
\bq
\yenv'' + \left[ 2 - \frac{3}{2} \Om(a) \right] \yenv' + \frac{1}{2} \Om(a) \left( \yenv^{-3}-1 \right) \yenv = 0 \,, \label{eq:yenv}
\eq
which follows from Eq.~(\ref{eq:shellmotion}) in the limit $\Delta\zeta/\zeta\rightarrow0$, or equivalently $\mathcal{F}\rightarrow0$.
We solve the system of differential equations~(\ref{eq:yhal}) and (\ref{eq:yenv}) with the initial conditions set in the matter-dominated regime,
\bq
y_{\rm h/env, i} = 1 - \frac{\delta_{\rm h/env, i}}{3} \,, \ \ \ \ \ y_{\rm h/env, i}' = - \frac{\delta_{\rm h/env, i}}{3} \,,
\eq
for an initial scale factor $a_{\rm i} \ll 1$.
We  define the effective linear overdensity
\bq
\delta_{\rm h/env}({\bf x}; \zeta_{\rm h/env}) \equiv \frac{D(a)}{D(a_{\rm i})} \delta_{\rm h/env, i} \,, \label{eq:extrapolation}
\eq
which is extrapolated from the initial overdensity to late times using the linear growth function of $\Lambda$CDM, $D(a)$.
In particular, we use Eq.~(\ref{eq:extrapolation}) to define the linear collapse and environmental densities, $\deltac^{f(R)}$ and $\denv$, respectively.
In practice, we evolve Eq.~(\ref{eq:yhal}) from $\delta_{\rm h, i}$ to the scale factor where it produces a singularity, to which we then use Eq.~(\ref{eq:extrapolation}) to linearly extrapolate $\delta_{\rm h/env, i}$ and define $\deltac^{f(R)}$ and $\delta_{\rm env}$. This effective approach evades complications from the scale-dependent growth in $f(R)$ gravity.


\begin{figure}[t]
\begin{center}
\includegraphics[width=0.6\columnwidth]{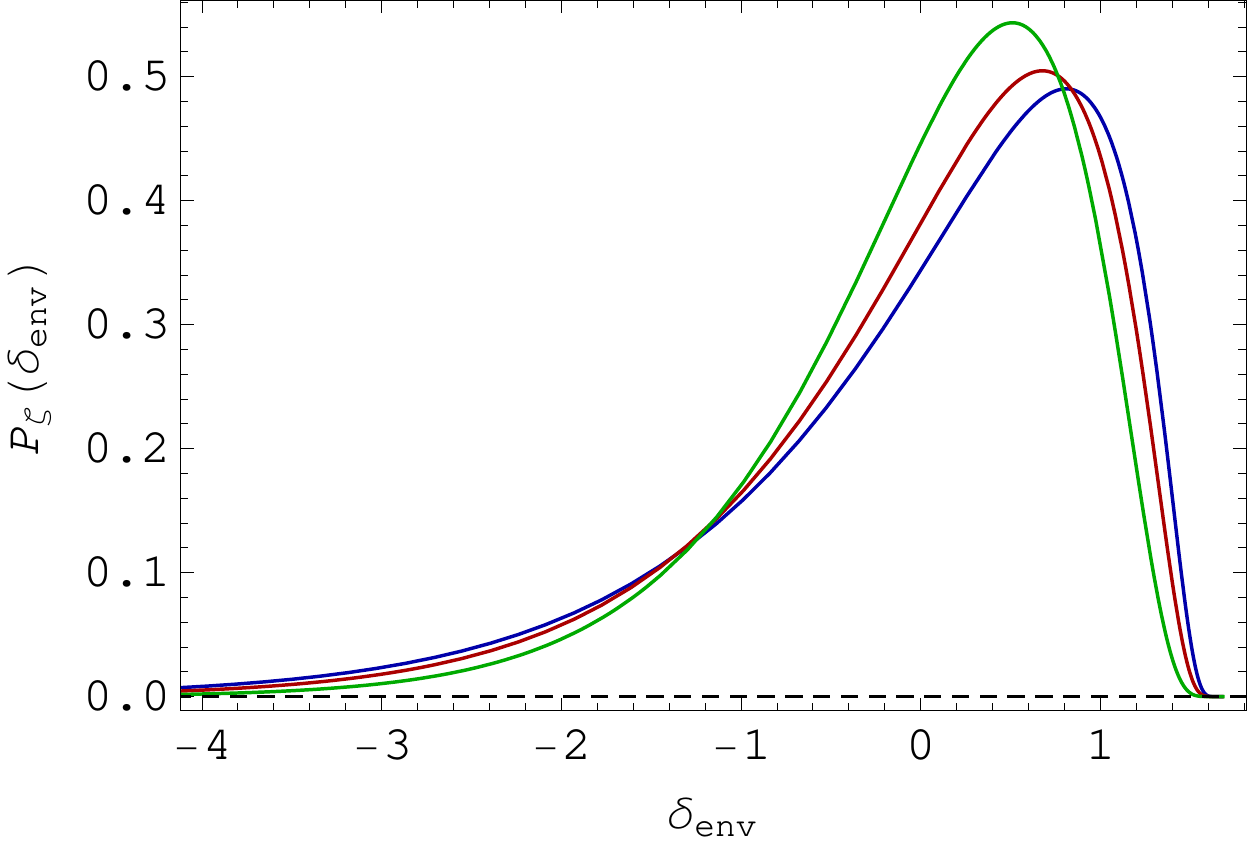}
\caption{Probability distributions of the Eulerian environment from Eq.~\ref{eq:eulerian} at three different redshifts, $z=0$ (blue), $z=0.2$ (red) and $z=0.5$ (green). In our spherical collapse calculations, at a given collapsing redshift we select the environmental density at the peak of the corresponding distribution.}
\label{fig:prob_env}
\end{center}
\end{figure}

As can be seen from Eq.~(\ref{eq:thshsphcoll}), the spherical collapse density, and therefore structure formation in chameleon $f(R)$ gravity, is dependent on both the mass of the halo formed, through $\RTH$, and its environmental density $\denv$ or $\delta_{\rm env,i}$.
To correctly reproduce the abundance of halos for a given mass measured in $N$-body simulations and to perform consistent tests of chameleon $f(R)$ gravity against observations, we determine the halo mass function averaged over the different environments.
Following~\cite{Li:2012ez,Lombriser:2013wta,Lombriser:2013eza}, we define the size of the environment as an Eulerian (physical) radius of $\zeta=5h^{-1}~{\rm Mpc}$ and approximate the probability distribution of an Eulerian environmental density $\denv$ as~\cite{Lam:2007qw,Lam:2012fa}
\bq
P_{\zeta}(\denv) = \frac{\beta^{\varpi/2}}{\sqrt{2\pi}} \left[ 1 + (\varpi - 1)\frac{\denv}{\deltacLCDM} \right] \left( 1 - \frac{\denv}{\deltacLCDM} \right)^{-\varpi/2-1} \exp \left[ -\frac{\beta^{\varpi}}{2} \frac{\denv^2}{(1 - \denv/\deltacLCDM)^{\varpi}} \right] \,, \label{eq:eulerian}
\eq
where $\beta = (\zeta/8)^{3/\deltacLCDM} / \sigma_8^{2/\varpi}$ with $\sigma_8$ being the present-day linear r.m.s.~density perturbation in spheres with radius $8h^{-1}$Mpc, $\deltacLCDM$ is the linearly extrapolated $\Lambda$CDM spherical collapse threshold, and $\varpi = \deltacLCDM \gamma$ with
\bq
\gamma = - \frac{\rmd \ln S_{\xi}}{\rmd \ln M_{\rm env}} = \frac{\tilde{n}_{\rm s}+3}{3} \,.
\eq
We use the Lagrangian (or initial comoving) radius $\xi=8h^{-1}~{\rm Mpc}$ with $S_{\xi}=\sigma_8^2$, $\tilde{n}_{\rm s}$ is the slope of the matter power spectrum on large scales at $a_{\rm i}\ll1$ in the matter era after turn over, and we assume that the environment evolves according to $\Lambda$CDM.

The distribution $P_{\zeta}(\denv)$ is shown in Fig.~\ref{fig:prob_env} for three different redshifts.
We will use the peak of the environmental distribution at a given redshift to approximate the environmentally averaged linear collapse density $\langle\deltac\rangle_{\rm env}$ and with that the observed average halo mass function.
More detailed discussions on alternative averaging procedures, comparisons between them, and further details on the role of the environment in chameleon modifications can be found in~\cite{Li:2012ez,Lam:2012fa,Lombriser:2013wta,Lombriser:2013eza}.


\subsection{Chameleon screening refinement}\label{sec:refinement}

In~\cite{Lombriser:2013wta,Lombriser:2013eza,Lombriser:2014dua} it was shown that at $z=0$ the spherical collapse model extended to incorporate a dependence on the environment gives a good description of the number density of virialized objects as a function of mass, i.e. of the halo mass function. However, for our purposes this approach is too simplistic, in that as described below it cannot capture in full detail the complex nonlinear dynamics of structure formation in $f(R)$ gravity. 

Due to the breakdown of Birkhoff's theorem spherical top-hat overdensities cannot be treated as close FRW universes, since their evolution also depends on the external matter distribution. Because of  this, an initially homogeneous spherical overdensity will evolve a profile resulting from the Yukawa-like fifth force in regions where the chameleon mechanism is not in action~\cite{Kopp:2013lea,Borisov:2011fu}. A possible solution to this problem consists in evolving the full set of field equations for an average initial density profile~\cite{Kopp:2013lea}.

On the other hand, the fact that dark matter halos possess higher-density internal substructures increases the chameleon efficiency in suppressing modifications of gravity~\cite{Li:2011uw}. Moreover, departures from the spherical collapse approximation in $f(R)$ gravity might also impact the growth of nonlinear structures to a greater extent than in GR. In fact, as it was first noticed in~\cite{JonesSmith:2011tn} and further investigated in~\cite{Burrage:2014daa}, the shape of extended objects affects the chameleon screening reducing its effectiveness with increasing ellipticity. This also introduces ``chameleonic'' torques that might have a measurable impact on the halo mass accretion history. In addition, the authors in~\cite{Pourhasan:2011sm} found that back-reaction effects can boost the chameleon efficiency in minor mergers depending on the mass of the infalling halo.

The complexity of the various physical processes involved, the extent of their individual contributions, as well as their interplay, make this problem amenable to semi-analytical modeling. In this work we opt for a correction of the spherical collapse predictions presented in Sec.~\ref{sec:sctheory} that is inspired by the phenomenological parameterized post-Friedmann (PPF) approach employed in Li \& Hu~(2011)~\cite{Li:2011uw}. Here, instead of applying this prescription to the variance of the linear density field while fixing the spherical collapse threshold to the $\Lambda$CDM value, we incorporate the PPF-inspired modifications through an effective collapse threshold $\dceff$, and use the same $\Lambda$CDM mass variance $\sigma(M)$ both for GR and $f(R)$ gravity. More specifically, for each background cosmology and collapse redshift $z_{\rm c}$ we define
\bq
\dceff(M,z_{\rm c}) \equiv \epsilon(M,z_{\rm c} | M_{\rm th}^{(1)},M_{\rm th}^{(2)},\eta,\vartheta,\chi) \times \deltac^{f(R)}(M,z_{\rm c},\denvpeak),
\label{eq:dceff}
\eq
where $\deltac^{f(R)}$ is evaluated following the method outlined in Sec.~\ref{sec:sctheory} at the environmental density where the distribution in Eq.~\eqref{eq:eulerian} peaks, $\denvpeak$. The correction factor is given by
\bq
\epsilon(M,z_{\rm c} | M_{\rm th}^{(1)},M_{\rm th}^{(2)},\eta,\vartheta,\chi) = \frac{1 + (M/M_{\rm th}^{(1)})^\eta(\deltacLCDM/\deltac^{f(R)})^\chi 
														+ (M/M_{\rm th}^{(2)})^\vartheta(\deltac^{f(R)}/\deltacLCDM)}
														{1 + (M/M_{\rm th}^{(1)})^\eta + (M/M_{\rm th}^{(2)})^\vartheta}.
\label{eq:correction}
\eq
The quantities $M_{\rm th}^{(1)}$, $M_{\rm th}^{(2)}$, $\eta$, $\vartheta$ and $\chi$ are free parameters that we will obtain by fitting our halo mass function model to the halo abundance measured from high-resolution cosmological simulations (see Secs.~\ref{sec:simulations} and~\ref{sec:hmf}). Before this, however, we can simplify the derivation of these parameters on the basis of theoretical and heuristic considerations. Similarly to Li \& Hu (2011)~\cite{Li:2011uw}, we consider $M_{\rm th}^{(1)}$ and $M_{\rm th}^{(2)}$ threshold masses controlling the transition between the $\deltac^{f(R)}$ and $\deltacLCDM$. 
As noted in~\cite{Lombriser:2014dua}, in the original PPF approach \cite{Li:2011uw} one can derive the scaling $M_{\rm th} \sim |f_{R0}|^{3/2}$ from Eq.~\eqref{eq:thshsphcoll}. Here, we apply this result to our threshold masses, and also include the dependence on $\Om$ and $n$. By interpreting $M_{\rm th}$ as the mass of an isolated halo with an interior scalaron profile approaching the minimum of the effective potential at its center, Eq.~\eqref{eq:thshsphcoll} implies
\bq
\frac{\Delta \zeta}{\zeta} = 1 \sim \frac{|f_{R0}|}{\Om M_{\rm th}^{2/3} \Delta_{\rm vir}^{1/3}(1+z_{\rm c})^{4+3n}} \left( \frac{1+4\frac{\Olam}{\Om}}{1+4\frac{\Olam}{\Om}(1+z_{\rm c})^{-3}} \right)^{n+1},
\label{eq:thickshllim}
\eq
where we have used $\tilde{\rho}_{\rm in}(z_{\rm c})\approx\Delta_{\rm vir}(\Om,z_{\rm c}) \gg 1$ and $\tilde{\rho}_{\rm out}=1$, with $\Delta_{\rm vir}$ denoting the virial overdensity in GR as a function of cosmology and collapse redshift. Assuming that we know the threshold mass $\tilde{M}_{\rm th}(\tilde{z}_{\rm c})$ for some particular set of parameters $\{ \tilde{\Omega}_{\rm m}, \tilde{f}_{R0}, \tilde{n} \}$ and redshift $\tilde{z}_{\rm c}$, we can then employ Eq.~\eqref{eq:thickshllim} to map this mass to any other combination of parameters as
\bqa
M_{\rm th} = &\tilde{M}_{\rm th}& \left( \frac{|f_{R0}|}{|\tilde{f}_{R0}|} \right)^{3/2} \left( \frac{\tilde{\Omega}_{\rm m}}{\Om} \right)^{3/2}
			\left( \frac{\tilde{\Delta}_{\rm vir}}{\Delta_{\rm vir}} \right)^{1/2} (1+\tilde{z}_{\rm c})^{-\frac{9}{2}(n-\tilde{n})} \nonumber \\
		&\times& \frac{\left[ 1+4\frac{\tilde{\Omega}_\Lambda}{\tilde{\Omega}_{\rm m}}(1+\tilde{z}_{\rm c})^{-3} \right]^{\frac{3}{2}(\tilde{n}+1)}}
				{\left[ 1+4\frac{\Olam}{\Om}(1+\tilde{z}_{\rm c})^{-3} \right]^{\frac{3}{2}(n+1)}}
				\frac{\left[ 1+4\frac{\Olam}{\Om} \right]^{\frac{3}{2}(n+1)}}
				{\left[ 1+4\frac{\tilde{\Omega}_\Lambda}{\tilde{\Omega}_{\rm}} \right]^{\frac{3}{2}(\tilde{n}+1)}}.
\label{eq:thmassfull}
\eqa
Note that Eq.~\eqref{eq:thmassfull} simply reduces to $M_{\rm th} \sim |f_{R0}|^{3/2}$ for $\Om = \tilde{\Omega}_{\rm m}$ and $n = \tilde{n}$. In this work we use $\tilde{\Omega}_{\rm m} = 0.281$, $|\tilde{f}_{R0}| = 10^{-5}$ and $\tilde{n} = 1$.

We do not impose any sign on $\eta,\vartheta$, and only require $M_{\rm th}^{(1)}, M_{\rm th}^{(2)} > 0$. Although we expect relatively small corrections to the spherical collapse solution, the domain over which the free parameters can change allows for rather generic deviations from the baseline $\deltac^{f(R)}$. These will push $\dceff$ either further away from or closer to the $\Lambda$CDM prediction. 
The remaining parameter $\chi$ controls how rapidly $\deltac^{\rm eff}$ approaches the $\Lambda$CDM threshold at high masses. This depends somewhat on the background scalaron field, and we found that the empirical relation
\bq
\chi = \frac{1}{2} - \frac{1}{5} \log_{10} \left( \frac{|f_{R0}|}{|\tilde{f}_{R0}|} \right)
\label{eq:chiaparam}
\eq
works well for our suites of simulations.

\begin{figure}[t]
\begin{center}
\includegraphics[width=0.6\columnwidth]{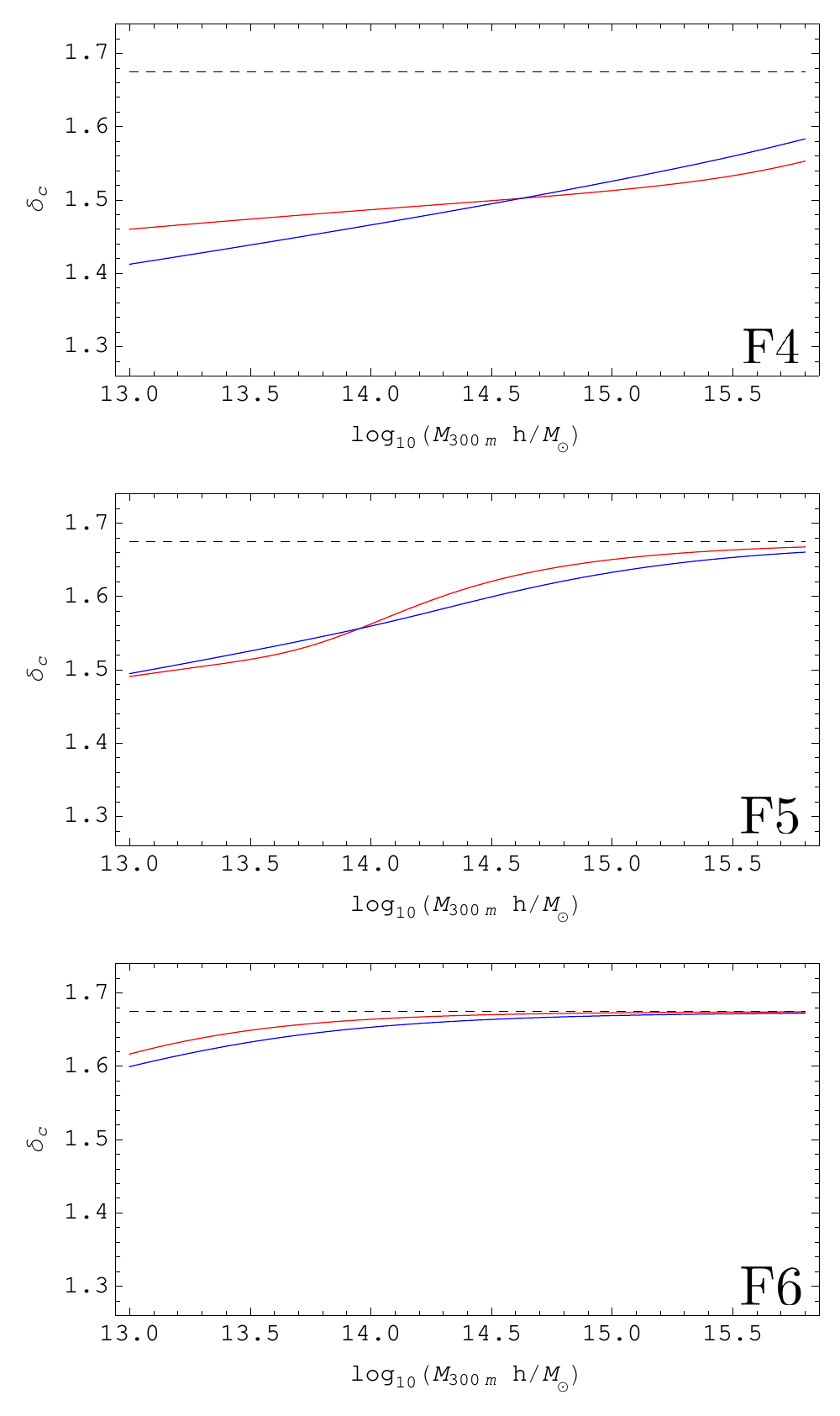}
\caption{Spherical collapse density thresholds $\deltac$ at $z=0$ and $\Om = 0.281$ for $|f_{R0}|=10^{-4},10^{-5},10^{-6}$ (F4, F5, F6 from top to bottom). In all panels, blue curves are obtained with the method described in Sec.~\ref{sec:sctheory}, and red curves correspond to the effective spherical collapse thresholds obtained after we correct them with Eq.~\eqref{eq:correction} calibrated with the high-resolution simulations of suite A in Table~\ref{tab:simulations}. Dashed lines mark the $\Lambda$CDM threshold. To avoid confusing notation, we define masses at an overdensity $\Delta = 300$ with respect to the background matter density for both $\deltac$ and $\deltac^{\rm eff}$. In reality, spherical collapse calculations are only meaningful for $\Delta = \Delta_{\rm vir}$.}
\label{fig:sc_predictions_Om028}
\end{center}
\end{figure}

In Fig.~\ref{fig:sc_predictions_Om028} we compare the spherical collapse predictions of Sec.~\ref{sec:sctheory} (blue lines) with the effective thresholds (red lines) from Eq.~\eqref{eq:dceff} that we calibrate using the suite A of high-resolution simulations listed in Table~\ref{tab:simulations}. As we explain in detail in Sec.~\ref{sec:hmf}, we incorporate Eq.~\eqref{eq:dceff} into the mass function model that we then fit to the halo abundances obtained from these simulations. For illustrative purposes, the effective thresholds shown in Fig.~\ref{fig:sc_predictions_Om028} correspond to those for the resulting best-fit values of the parameters in Eq.~\eqref{eq:dceff}. Despite the visible differences, the corrected, effective thresholds remain within a few percent from the original spherical collapse thresholds, which justifies our approach of introducing higher-order corrections. In principle, these effective quantities could be seen as averaged solutions to Eqs.~\eqref{eq:thshsphcoll}-\eqref{eq:yenv} over a suitable, yet unknown, environmental density distribution different from that of Eq.~\eqref{eq:eulerian}. Here, however, we refrain from giving any profound physical interpretation to such deviations, and remark that they can also be partly attributed to the difference between the virial overdensity $\Delta_{\rm vir}$ (dependent on redshift, mass and cosmology) and the fixed overdensity at which we define dark matter halos in our study (see Sec.~\ref{sec:hmf})~\cite{Schmidt:2008tn,Despali:2015yla}. Nevertheless, they hint to the possibility that the initial density profile, and the geometry and substructure of dark matter halos might leave a mass-dependent imprint on the averaged mass function unaccounted for in the spherical collapse treatment of Lombriser et al.~(2013)~\cite{Lombriser:2013wta}.

\section{Simulations} \label{sec:simulations}

The simulations used to calibrate the theoretical HMFs in this work were run with the {\sc ecosmog} code \cite{Li:2011vk}, which is an extension to the publicly available {\sc ramses} N-body and hydro code \cite{Teyssier:2001cp} for cosmological simulations in modified gravity theories. The code employs the particle-mesh technique with adaptive mesh refinement to compute the (modified) gravitational force. In short, it starts with what it is called a domain grid which is a regular mesh with $N_{\rm cell}^3 = 1024^3$ cells covering the cubic simulation box of size $L_{\rm box}$ (expressed in units of $h^{-1}$Mpc). A number of $N_p^3$ particles are evolved on this mesh from an initial redshift $z_{\rm ini}$. The density field on the mesh is obtained by a cloud-in-cell (CIC) or triangular-shaped cloud (TSC) interpolation to its cells; this is then used to compute the gravitational forces at the cell centres, which are next used to move the particles. To achieve higher force resolution in high-density regions, the code adaptively refines a cell if the effective particle number inside it exceeds some pre-defined criterion $N_{\rm ref}$ -- this has proved to be critical to ensure accuracy when solving the modified gravity force, which has a smaller amplitude in these regions due to the chameleon screening. The code is efficiently parallelized using {\sc mpi}, with domain decomposition achieved through a standard Peano-Hilbert space-filling curve. For more details on the {\sc ecosmog} code please refer to the original code paper for $f(R)$ gravity \cite{Li:2011vk} and its subsequent extensions to other models \cite{Brax:2012sd,Brax:2013ch,Li:2013dgp,Barreira:2013cg,Li:2013qg}.

Two suites of simulations are used in this work. Suite A was run using a flat cosmology with WMAP 9-year best-fit parameters
\begin{equation}
(\Omega_{\rm m}, \Omega_\Lambda, h, n_s, \sigma_8) = (0.281, 0.719, 0.697, 0.971, 0.82)\,,
\label{eq:suiteA_cosmo}
\end{equation}
where $n_s$ is the scalar spectral index of the primordial power spectrum. We also use suite B (see details in \cite{Li:2013pk}) with older, WMAP 3-year best-fit parameters
\begin{equation}
(\Omega_{\rm m}, \Omega_\Lambda, h, n_s, \sigma_8) = (0.240, 0.760, 0.730, 0.958, 0.80)\,.
\label{eq:suiteB_cosmo}
\end{equation}
As described in Sec.~\ref{sec:hmf}, we use suite A for the actual calibration of the theoretical HMF, and then compare this with simulation results from suite B to test how our fit works for this other cosmology.

In both suites, we simulated three variants of Hu-Sawicki $f(R)$ gravity with $n=1$ and different values of $f_{R0}$: $-10^{-6}$ (hereafter dubbed F6), $-10^{-5}$ (F5) and $-10^{-4}$ (F4). To quantify the modified gravity effects and compare them with our model of the ratio of $f(R)$ to GR, we also run a GR case for each suite. All simulations within a suite or realization started from the same initial redshift $z_{\rm ini}=49.0$ and were evolved until today ($z=0$). The initial conditions were generated using the publicly available {\sc mpgrafic} code \cite{Prunet:2008za}, which employs the standard Zel'dovich approximation to calculate the initial particle displacement and velocity fields. We used the same initial conditions for GR and $f(R)$ simulations within the same suite because at $z=49$ the effect of this modified gravity model on the particle distributions is negligible. Note that the $\sigma_8$ values quoted above are the $z=0$ linear-theory results for GR (they would be different for different variants of $f(R)$ models), and as such they should be considered as a characterisation of the initial conditions rather than of the large-scale matter clustering today. Further simulation specifications are summarised in Table~\ref{tab:simulations}.

\ctable[
    caption = {Specifications of the N-body simulations used in this work. $N_{\rm ref}$ -- the refinement criterion used in these simulations -- is 8.0 for suite A and 9.0 for suite B, and $m_p$ is the simulation particle mass.},
    label   = {tab:simulations},
pos =t,
mincapwidth=\textwidth,
]{ >{\centering\arraybackslash} m{0.6cm}  >{\centering\arraybackslash} m{2cm}  >{\centering\arraybackslash} m{1cm}  >{\centering\arraybackslash} m{2cm}  >{\centering\arraybackslash} m{2cm}  >{\centering\arraybackslash} m{3cm}  >{\centering\arraybackslash} m{2cm} }{}{                                                          				
\hline\hline
Suite & $L_{\rm box}$ & $N_p^3$ & density \mbox{interpolation} & force resolution& $m_p$ & number of realizations \\
\hline
A & $1024h^{-1}$Mpc & $1024^3$ & CIC &  31.2$h^{-1}$kpc & $7.8\times10^{10}h^{-1}M_\odot$ & $1$ \\
B & $1500h^{-1}$Mpc & $1024^3$ & TSC & 45.8$h^{-1}$kpc & $2.1\times10^{11}h^{-1}M_\odot$ & $6$ \\
\hline
\hline
}


\section{Halo mass function} \label{sec:hmf}

In this section we present our main results on the modeling of the $f(R)$ gravity halo mass function. Compared to previous works, we devote particular care to the mass binning used to derive halo abundances from simulations (see Sec.~\ref{sec:binnedhmf}), as well as to estimate the corresponding uncertainties. We also use our results to forecast approximate constraints from cluster number count data (see Sec.~\ref{sec:forecast}).

\subsection{Binned mass function from simulations}\label{sec:binnedhmf}

We identify dark matter halos in our simulations using the {\sc rockstar} halo finder~\cite{Behroozi:2011ju}, which by default obtains spherical overdensity (SO) masses from initial friends-of-friends (FOF) groups neglecting unbound particles. However, masses are defined observationally within spherical apertures of arbitrary size enclosing an overdensity that might not be entirely virialized. Since our goal is to have mass function predictions calibrated for X-ray, Sunyaev-Zel'dovich (SZ) and optical cluster surveys, we enable {\sc rockstar} to calculate strict SO masses including unbound particles, as well as particles that may reside outside of the FOF group associated with the halo. Here, we choose an average overdensity $\Delta=300$ such that the mass inside a sphere of radius $r_\Delta$ is
\bq
M_{300 {\rm m}} = \frac{4}{3}\pi r_\Delta^3 \rhomb(z)\Delta.
\eq
Winther et al.~(2015)~\cite{Winther:2015wla} showed that, even for $\Lambda$CDM, different modified gravity $N$-body codes produce mass functions that differ by as much as 10\%. It was also noticed that these discrepancies are approximately independent of the particular value of $f_{R0}$. Thus, taking the ratio of the HMF in $f(R)$ to that in GR reduces this scatter to a more competitive 1-4\% (for $z \lesssim 0.5$). In addition, the effects of baryonic physics on the $f(R)$ mass function can be potentially neglected when considering instead the ratio of this HMF to that of GR~\cite{Cataneo:2014kaa,Puchwein:2013lza}. For this, departures from pure dark matter predictions could be incorporated with the same level of accuracy through a baseline GR mass function calibrated with hydrodynamical simulations (see e.g.~\cite{Bocquet:2015pva}). Initially, one might think that mass function ratios may exhibit larger uncertainties compared to those of the individual mass functions. Nonetheless, halo abundances in GR and $f(R)$ are presumably strongly correlated, and it is reasonable to expect that on average such correlation reduces the errors on the ratios to the level of those on the corresponding mass functions (see Eq. (91) in~\cite{Brax:2012sd}). For these reasons, we consider the HMF ratios
\bq
\mathcal{R}_i^{\rm sim} \equiv \left< \mathcal{R}_i \right>_{\rm JK} = \left< \frac{N^{f(R)}_{h,i}}{N^{\rm GR}_{h,i}} \right>_{\rm JK}
\label{eq:sim_ratios}
\eq
our fundamental observables from the simulations. In Eq.~\eqref{eq:sim_ratios} $\left< \cdot \right>_{\rm JK}$ denotes the jackknife average, and $N^{f(R)}_{h,i}$ and $N^{\rm GR}_{h,i}$ are the number of halos in the $i$-th mass bin for $f(R)$ and GR, respectively. Also, we implicitly used the fact that volume and mass bin size are identical for the particular pair of simulations examined. We employ the unbiased jackknife average
\bq
 \left< \mathcal{R}_i \right>_{\rm JK} = N_{\rm JK}\overline{\mathcal{R}_i} - (N_{\rm JK} -1) \overline{\mathcal{R}_i^{\rm JK}},
 \label{eq:jk_average}
\eq
where $N_{\rm JK}$ is the number of simulation subvolumes, $\overline{\mathcal{R}}_i$ is the standard sample mean over the $N_{\rm JK}$ subvolumes, and the resampled jackknife estimates are
\bq
\overline{\mathcal{R}_i^{\rm JK}} = \frac{1}{N_{\rm JK}}\sum_{j =1}^{N_{\rm JK}} \mathcal{R}_{i,j}^{\rm JK}\,, 
\eq
with 
\bq
\mathcal{R}_{i,j}^{\rm JK} = \frac{1}{N_{\rm JK} -1}\sum_{k \neq j} \mathcal{R}_{i,k}\,. 
\label{eq:jk_resampling}
\eq
For our suite A of simulations in Table~\ref{tab:simulations} we divide each box in octants and remove one octant at a time from the full simulation volume to compute Eq.~\eqref{eq:jk_resampling}. We proceed similarly for our suite B, although in this case each jackknife subvolume corresponds to a different realization. 

Following Tinker et al.~(2008)~\cite{Tinker:2008ff}, and supported by results in~\cite{Crocce:2009mg,Hoffmann:2015mma}, we also adopt the jackknife method to estimate the error contributions on Eq.~\eqref{eq:sim_ratios} from both sample variance and Poisson noise. For each mass bin $i$ we have
\bq
\sigma_{\mathcal{R},i} = \sqrt{N_{\rm JK} - 1} \, s_{\mathcal{R}^{\rm JK},i}\,,
\label{eq:jk_error}
\eq
where the jackknife sample variance is
\bq
s_{\mathcal{R}^{\rm JK},i}^2 = \overline{(\mathcal{R}_i^{\rm JK})^2} - \left( \overline{\mathcal{R}_i^{\rm JK}} \right)^2.
\label{eq:jk_samplevariance}
\eq
On top of this error, we should also include the 1-4\% scatter found between $N$-body codes in~\cite{Winther:2015wla}, as well as the error introduced by assigning the HMF ratio in each mass bin to the center of the bin~\cite{Lukic:2007fc}. The former is comparable to the error from Eq.~\eqref{eq:jk_error} for masses up to $M \sim 10^{14.5} \Msunh$, and dominates over the bin center error in the same range for mass bins $\Delta\log_{10} M = 0.15$. For larger masses the contrary is true. For simplicity, however, we neglect both of these contributions since adding them would not considerably alter our best fits, and a full statistical analysis of the new HMF parameters is not within the scope of this paper. Note also that for the mass range of interest here, $10^{13} {\rm -} 10^{15.5} \Msunh$, we adopt a bin size for which we expect our results of the HMF ratios to be converged within the errors (see e.g.~\cite{Bhattacharya:2010wy,Hoffmann:2015mma}). Furthermore, previous works showed that in this mass range bins are mostly uncorrelated~\cite{Smith:2011vm,Hoffmann:2015mma}, with none or very marginal impact on the best fitting values of the model parameters~\cite{Hoffmann:2015mma}. Hence, in what follows we ignore all covariances between mass bins, the effect of which should be negligible for our results.

Mass function ratios are also suitable to alleviate the consequences of other numerical inaccuracies. Numerical transients related to Zel'dovich initial conditions (1LPT) are responsible for a deficit of massive halos compared to results obtained from second order initial conditions (2LPT)~\cite{Crocce:2009mg,Reed:2012ih}. However, assuming that the same correction applies to the HMF's of both $f(R)$ and GR obtained with 1LPT at $z_{\rm ini} = 49$, for final redshifts in the range $z_{\rm fin} \in [0,0.5]$ we estimate a conservative average difference between 1LPT and 2LPT HMF ratios of 1\%, which is well within our jackknife errors~\cite{Taruya:2016jdt}. As a final note, Knebe et al.~(2013)~\cite{Knebe:2013xwz} found a 10\% scatter among mass functions derived using different halo finders. Also in this case, HMF ratios are expected to contain these differences safely within our estimates from Eq.~\eqref{eq:jk_error}. 

\subsection{Modeling and fits}\label{sec:fits}

We derive our predictions for the ratios of the mass function in $f(R)$ over the mass function in GR from the prescription given by Sheth \& Tormen~(1999)~\cite{Sheth:1999mn}. In this framework, the comoving number density of halos in a logarithmic mass bin around a mass $M$ is
\bq
n_{\ln M} \equiv \frac{dn}{d\ln M} = \frac{\rhomb}{M}\frac{d\ln \nu}{d\ln M}\nu f(\nu),
\label{eq:STmfcn}
\eq
where $\nu = \deltac/\sigma(M,z)$ is the peak height, with 
\bq
\sigma^2(M,z) = \int \frac{d^3 k}{(2\pi^3)} |\tilde{W}_R(k)|^2 P_L(k,z).
\eq
Here, $P_L(k,z)$ is the linear power spectrum\footnote{For the linear calculations of both GR and $f(R)$ we evaluate $\Lambda$CDM matter power spectra (see more details later on in the text) using the publicly available code \sc{camb}~\cite{Lewis:1999bs}.} at redshift $z$ and $\tilde{W}_R(k)$ is the Fourier transform of the top hat window function of radius $R$ that encloses a mass $M = 4\pi R^3 \bar{\rho}_m/3$. We use $\deltac = \deltacLCDM(z)$ for GR, which we evaluate as~\cite{Nakamura:1996tk}
\bq
\deltacLCDM(z) \approx  \frac{3}{20}(12\pi)^{2/3}\left[ 1 + 0.0123\log_{10}\Om(z) \right],
\eq
and $\deltac = \deltac^{\rm eff}(M,z)$ given in Eq.~\eqref{eq:dceff} for $f(R)$. The Sheth-Tormen (ST) multiplicity function in Eq.~\eqref{eq:STmfcn} is parametrized as
\bq
\nu f(\nu) = A\sqrt{\frac{2}{\pi}a\nu^2}\left[ 1 + (a\nu^2)^{-p} \right] \exp \left[ -a\nu^2/2 \right],
\label{eq:nufnu}
\eq
where $(a,p,A)$ are free parameters defining the high-mass cutoff, the low-mass shape and the normalization of the mass function, respectively. For these, we employ the recent fits from Despali et al.~(2015)~\cite{Despali:2015yla}, where they extended the previous ST fits to be function of a generic overdensity $\Delta$. For easy reference, we report these results here,
\bq
\begin{aligned}
\label{eq:STfits}
a &= 0.4332x^2 + 0.2263x + 0.7665, \\
p &= -0.1151x^2 + 0.2554x + 0.2488,  \\ 
A &= -0.1362x + 0.3292\,,
\end{aligned}
\eq
where $x=\log_{10}[\Delta/\Delta_{\rm vir}(z)]$. In our approach all the modifications of gravity are encoded in $\deltac^{\rm eff}$, thus Eq.~\eqref{eq:STfits} is used both in GR and $f(R)$, and we approximate the virial overdensity as~\cite{Bryan:1997dn}
\bq
\Delta_{\rm vir}(z)  = \frac{18\pi^2 - 82\left[ 1-\Om(z) \right] - 39\left[ 1-\Om(z) \right]^2}{\Om(z)}.
\label{eq:virialOD}
\eq
A similar argument applies to the mass variance of the linear density field. In GR the statistics of collapsed objects at any redshift is fully determined by the initial linear density field 
\bq
\nu_{\rm ini} \equiv \frac{\delta_{\rm i}(z_{\rm c})}{\sigma(M,z_{\rm i})} = \frac{D(z_{\rm c})\delta_{\rm i}(z_{\rm c})}{D(z_{\rm c})\sigma(M,z_{\rm i})}
													= \frac{\deltacLCDM(z_{\rm c})}{\sigma(M,z_{\rm c})} = \nu(z_{\rm c}),
\label{eq:inipeak}
\eq
where $\delta_{\rm i}(z_{\rm c})$ represents the density contrast at an initial redshift $z_{\rm i}$ that will eventually produce a halo at a formation time $z_{\rm c}$. Considering that the initial conditions are set such that $\sigma_{f(R)}(M,z_{\rm i}) = \sigma_{\rm GR}(M,z_{\rm i})$ for all scales of interest, then enforcing Eq.~\eqref{eq:inipeak} also in $f(R)$ effectively implies $\sigma_{f(R)}(M,z) = \sigma_{\rm GR}(M,z)$ at all redshifts~\cite{Kopp:2013lea,Lombriser:2013wta}.

We define our theoretical mass function ratios using Eq.~\eqref{eq:STmfcn} together with Eq.~\eqref{eq:nufnu} as
\bq
\mathcal{R}^{\rm theo}(M) \equiv \frac{n_{\ln M} |_{\scriptsize f(R)}}{n_{\ln M} |_{\rm GR}},
\label{eq:theoratios}
\eq
which depend on the set of free parameters $M_{\rm th}^{(1)}$, $M_{\rm th}^{(2)}$, $\eta$ and $\vartheta$ introduced in Sec.~\ref{sec:refinement}. For our fitting analysis, we employ the suite A of high-resolution simulations in Table~\ref{tab:simulations}, and consider each redshift snapshot $z_{\rm c} \in \left\{ 0.0,0.1,0.2,0.3,0.4,0.5 \right\}$ separately. We obtain the best-fit values by minimizing
\bq
\chi^2(M_{\rm th}^{(1)},M_{\rm th}^{(2)},\eta,\vartheta) = \sum_{i}\frac{[\mathcal{R}_i^{\rm sim} - \mathcal{R}^{\rm theo}(M_i)]^2}{\sigma_{\mathcal{R},i}^2},
\eq
where the sum is over mass bins with at least 20 halos to limit the effect of Poisson noise at high masses, and $M_i$ denotes the mass at the bin center. We first fit the F5 simulations to find $\tilde{M}_{\rm th}^{(1)}(z_{\rm c})$ and $\tilde{M}_{\rm th}^{(2)}(z_{\rm c})$, which we then rescale to the other values of $f_{R0}$ using Eq.~\eqref{eq:thmassfull}. Hence, in the F4 and F6 cases we only fit for $\eta$ and $\vartheta$. Below, we provide fitting functions for the relevant free parameters entering Eq.~\eqref{eq:correction}. To achieve enough flexibility without including a large number of terms, for $\eta$ and $\vartheta$ we opted for 2-dimensional surfaces described by cubic polynomials in redshifts with coefficients depending quadratically on $\log_{10}|f_{R0}|$:
\bq
\begin{aligned}
\label{eq:HMFzfits}
\tilde{M}_{\rm th}^{(1)}(z) &= 13.8528 - 0.5981 z - 2.7073 z^2 + 4.1907 z^3, \\
\tilde{M}_{\rm th}^{(2)}(z) &= 13.9720 - 0.9003 z - 2.9086 z^2 + 5.4463 z^3,  \\ 
\eta(f_{R0},z) &= \eta_0(f_{R0}) + \eta_1(f_{R0}) z + \eta_2(f_{R0}) z^2 + \eta_3(f_{R0}) z^3, \\
\vartheta(f_{R0},z) &= \vartheta_0(f_{R0}) + \vartheta_1(f_{R0}) z + \vartheta_2(f_{R0}) z^2 + \vartheta_3(f_{R0}) z^3, 
\end{aligned}
\eq
where the $\eta_i$ coefficients are
\bq
\begin{aligned}
\label{eq:etafits}
\eta_0(f_{R0}) &= -46.1022 - 18.5382 \log_{10}|f_{R0}| - 1.7648 \left( \log_{10}|f_{R0}| \right)^2, \\
\eta_1(f_{R0}) &= 6.0520 + 4.8043 \log_{10}|f_{R0}| + 0.7146 \left( \log_{10}|f_{R0}| \right)^2,  \\ 
\eta_2(f_{R0}) &= -398.9787 - 180.5379 \log_{10}|f_{R0}| - 19.8292 \left( \log_{10}|f_{R0}| \right)^2, \\
\eta_3(f_{R0}) &= 429.3937 + 201.2807 \log_{10}|f_{R0}| + 22.9045 \left( \log_{10}|f_{R0}| \right)^2, 
\end{aligned}
\eq
and the $\vartheta_i$ coefficients are
\bq
\begin{aligned}
\label{eq:thetafits}
\vartheta_0(f_{R0}) &= -19.6362 - 8.1120 \log_{10}|f_{R0}| - 0.7744 \left( \log_{10}|f_{R0}| \right)^2, \\
\vartheta_1(f_{R0}) &= 67.6699 + 25.9151 \log_{10}|f_{R0}| + 2.4720 \left( \log_{10}|f_{R0}| \right)^2,  \\ 
\vartheta_2(f_{R0}) &= -651.2764 - 274.0971 \log_{10}|f_{R0}| - 28.4491 \left( \log_{10}|f_{R0}| \right)^2, \\
\vartheta_3(f_{R0}) &= 726.4060 + 311.8720 \log_{10}|f_{R0}| + 33.1439 \left( \log_{10}|f_{R0}| \right)^2. 
\end{aligned}
\eq
Note that all the expressions above are only valid in the redshift range $0\leqslant z \leqslant0.5$, for $10^{-6} \leqslant |f_{R0}| \leqslant 10^{-4}$ and $\Delta = 300$. 

\begin{figure}[t]
\begin{center}
\includegraphics[width=0.48\columnwidth]{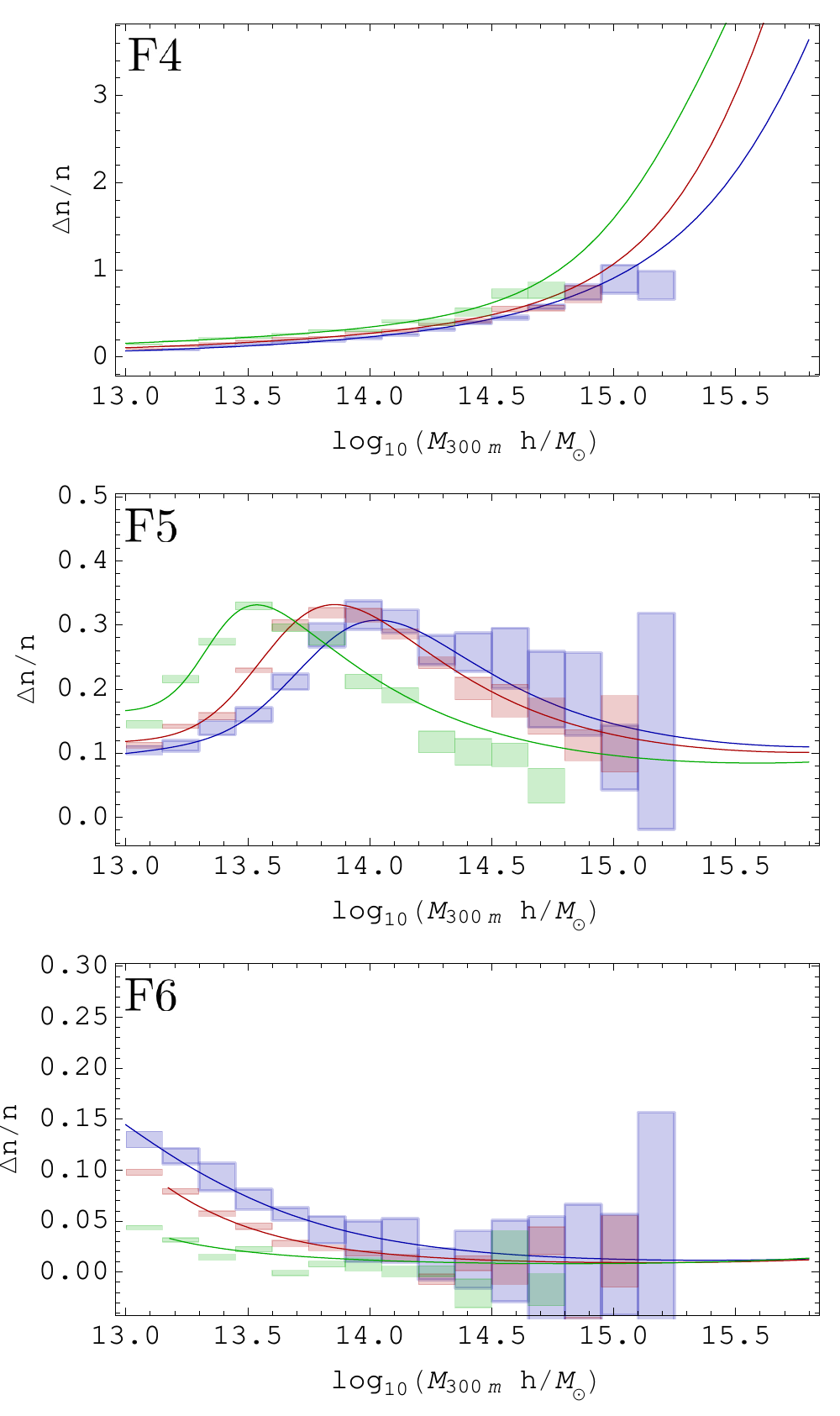}
\quad
\includegraphics[width=0.48\columnwidth]{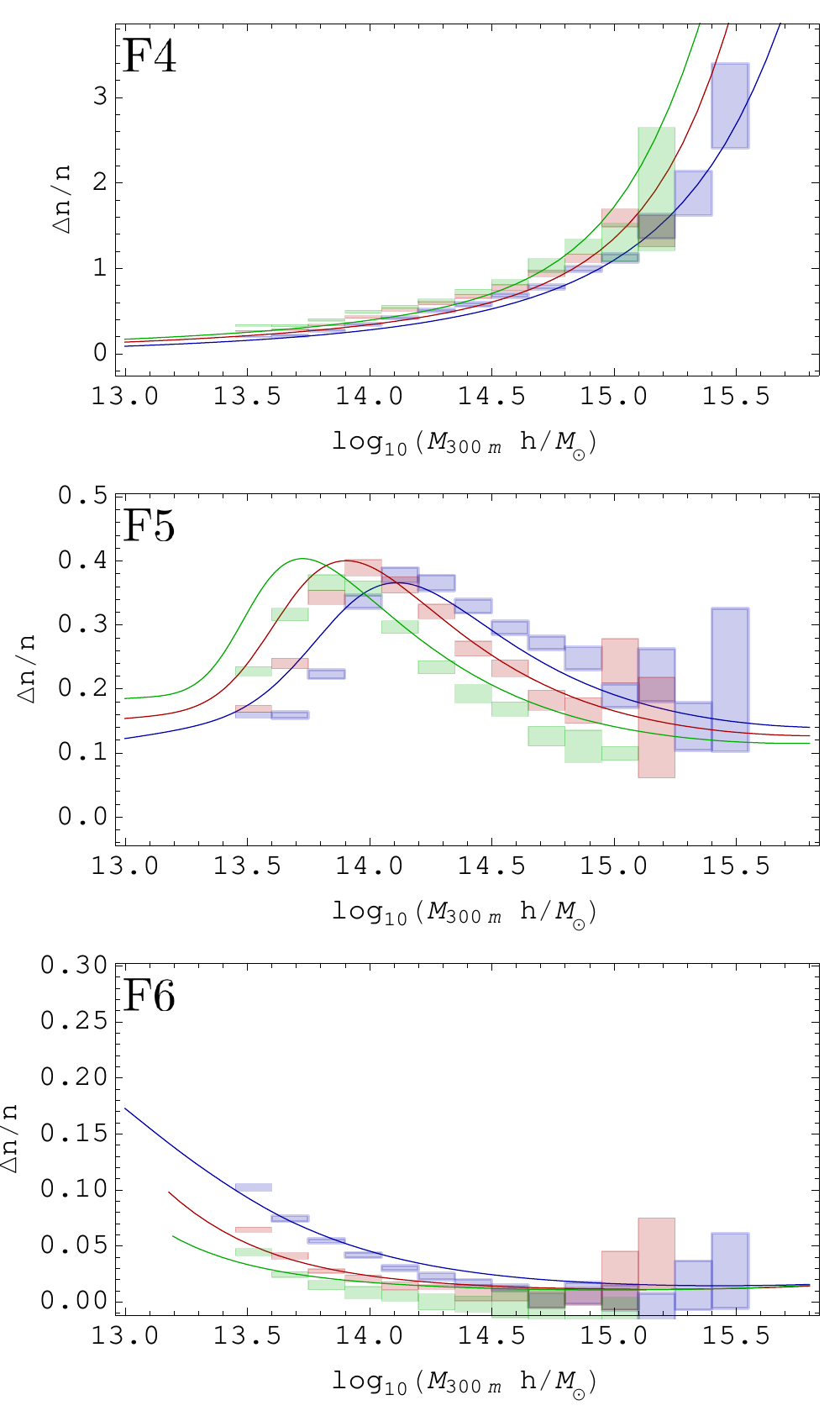}
\end{center}
\caption{Mass function fractional enhancements in $f(R)$ relative to GR as a function of redshift and background scalaron amplitude (from top to bottom, $|f_{R0}| = 10^{-4}, 10^{-5}, 10^{-6}$). {\it Left}: comparison between our fits (lines) and halo abundance bins (rectangles) from the high-resolution simulations of suite A, for $z=0$ (blue), $z=0.2$ (red) and $z=0.5$ (green). {\it Right:} the same as in the left panel but for the lower-resolution simulations of suite B, and for $z=0$ (blue), $z=0.25$ (red) and $z=0.44$ (green). We find our fits to be within 5\% precision for $M \gtrsim 10^{14} \Msunh$ (see also main text for further details).}
\label{fig:HMF_predictions}
\end{figure}

\begin{figure}[t]
\begin{center}
\includegraphics[width=0.48\columnwidth]{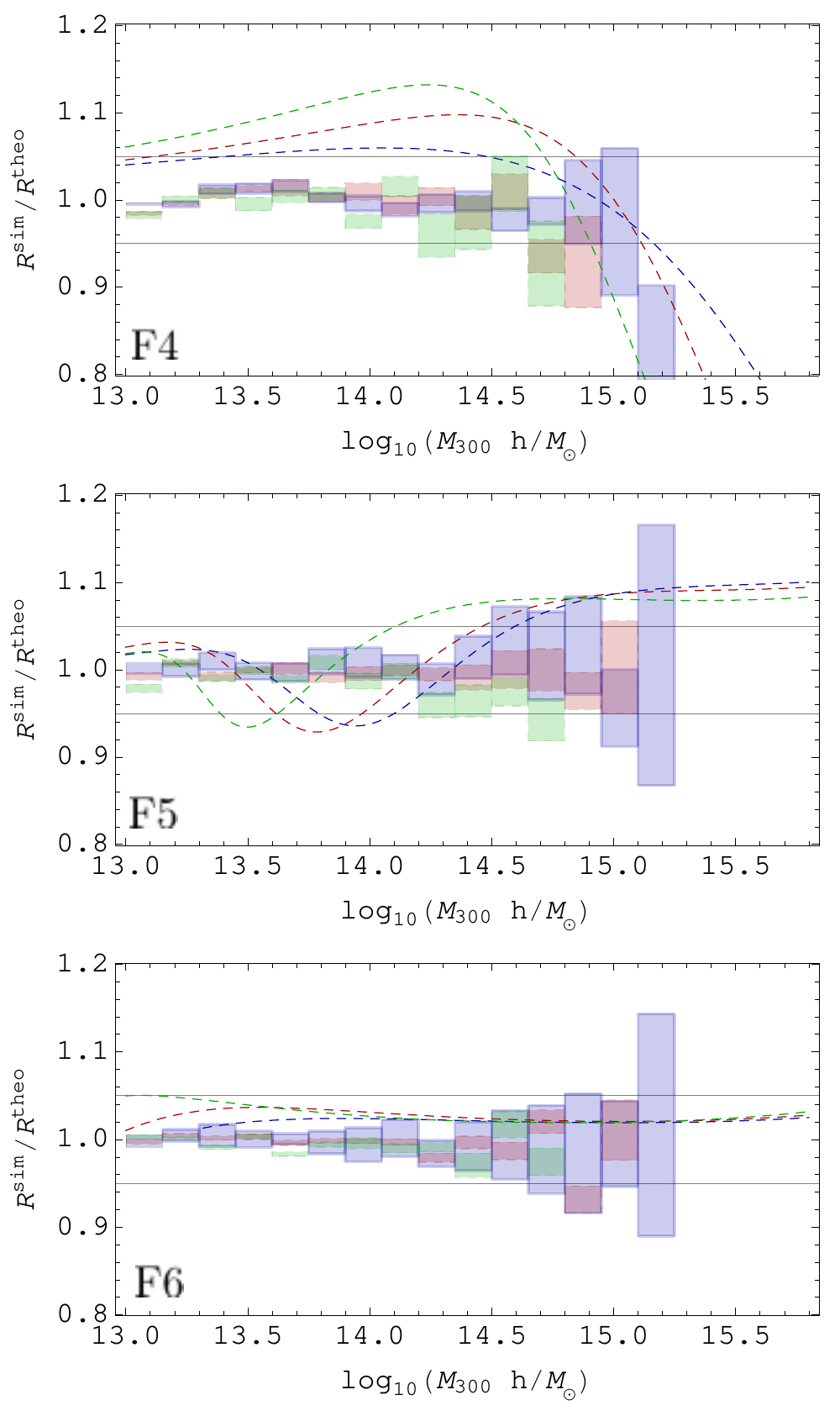}
\end{center}
\caption{Ratios of the measured $f(R)$ mass function enhancements from the high-resolution simulations of suite A with respect to the same quantity from the fitting functions Eqs.~\eqref{eq:HMFzfits}-\eqref{eq:thetafits} for three selected redshifts, $z=0$ (blue), $z=0.2$ (red) and $z=0.5$ (green). The background scalaron amplitude varies across the three panels (from top to bottom, $|f_{R0}| = 10^{-4}, 10^{-5}, 10^{-6}$), and horizontal lines mark $\pm 5\%$ deviations from our predictions. For comparison, we also show the effect of neglecting the correction factor Eq.~\eqref{eq:correction} and using $\deltac^{f(R)}$ alone (dashed lines).}
\label{fig:simtheo_reldiff}
\end{figure}

Based on Eqs.~\eqref{eq:HMFzfits}-\eqref{eq:thetafits}, the left panel of Fig.~\ref{fig:HMF_predictions} shows our predictions for the $f(R)$ to GR HMF ratios (lines), and how these compare to the corresponding ratios measured from the simulations in mass bins using Eq.~\eqref{eq:sim_ratios} (rectangles). Our fits perform very well for the three selected redshifts ($z=0$, in blue; $z=0.2$, in red; $z=0.5$, in green) and for the three representative background field values (F4, F5 and F6), with deviations of $\lesssim 5\%$ from the mean ratios over the entire mass range (see also Fig.~\ref{fig:simtheo_reldiff}). Assuming that $\eta$ and $\vartheta$ are slowly varying functions of the cosmological parameters (in particular $\Om$), we can also employ Eqs.~\eqref{eq:HMFzfits}-\eqref{eq:thetafits} along with Eq.~\eqref{eq:thmassfull} to predict the HMF ratios for other background cosmologies. The simulations of suite B in Table~\ref{tab:simulations} were run for a sufficiently different background cosmology from the one we used to calibrate our relations (cfr. Eq.~\eqref{eq:suiteA_cosmo} and Eq.~\eqref{eq:suiteB_cosmo}) to provide a good test bench in which to assess the validity of these results for other background cosmologies. In the right panel of Fig.~\ref{fig:HMF_predictions}, we illustrate the predictive power of our fits for the F4, F5 and F6 $f(R)$ cosmologies in suite B (top to bottom panels), as well as for snapshots at redshift $z=0$ (blue), and two other redshifts somewhat different from those in suite A, $z=0.25$ (red) and $z=0.44$ (green). Once again, the agreement with simulations is very good for $M \gtrsim 10^{14} \Msunh$, although discrepancies are visible for smaller masses. Obviously, one possible reason for such behavior is a lack of `universality' in our fitting parameters, especially for quite distinct cosmologies, such as WMAP3 and WMAP9. However, another plausible explanation is that the lower mass and force resolutions in suite B, together with a different density interpolation method and a less stringent refinement criterion, might affect the low-mass halo abundance in GR and $f(R)$ differently. 

Indeed, particle-mesh codes with coarse domain grids suppress the abundance of low-mass halos~\cite{O'Shea:2003gd,Lukic:2007fc}. As mentioned in Sec.~\ref{sec:simulations}, {\sc ecosmog} employs adaptive mesh refinement to improve force resolution, with the refinement criterion being a rather important parameter for the code performance. O'Shea et al.~(2005)~\cite{O'Shea:2003gd} recommend a domain grid twice as fine as the mean interparticle spacing, as well as a low refinement threshold to achieve enough force resolution and capture small density peaks at early times. Although the size of our domain grid cells $\Delta_{\rm cell} = L_{\rm box}/N_{\rm cell}$ is equal to or larger than the mean interparticle separation $\Delta_p = L_{\rm box}/N_p$, the low refinement threshold $N_{\rm ref}$ might help to reach an effective domain grid twice as fine as the original one, $\Delta_{\rm cell}' \approx \Delta_{\rm cell}/2$. Luki\'{c} et al.~(2007)~\cite{Lukic:2007fc} proposed a criterion for the minimum number of particles required to accurately resolve a halo that at $z=0$ reads
\bq
N_p^{\rm halo} \gtrsim 4.2 \left( \frac{\Delta_{\rm cell}'}{\Delta_p} \right)^3 \frac{\Delta}{\Om}.
\label{eq:nhalomin}
\eq
Recalling that for this work $\Delta=300$, and using the information in Table~\ref{tab:simulations} together with Eqs.~\eqref{eq:suiteA_cosmo} and~\eqref{eq:suiteB_cosmo}, we have that $M_{\rm halo}^{\rm min} \approx 10^{13.5} \Msunh$ and $M_{\rm halo}^{\rm min} \approx 10^{14} \Msunh$ for suites A and B respectively. Interestingly, this is consistent with our findings in Fig.~\ref{fig:HMF_predictions} for the lower-resolution simulations of suite B. 

\subsection{Approximate forecasts}\label{sec:forecast}

\begin{figure}[t]
\begin{center}
\includegraphics[width=0.48\columnwidth]{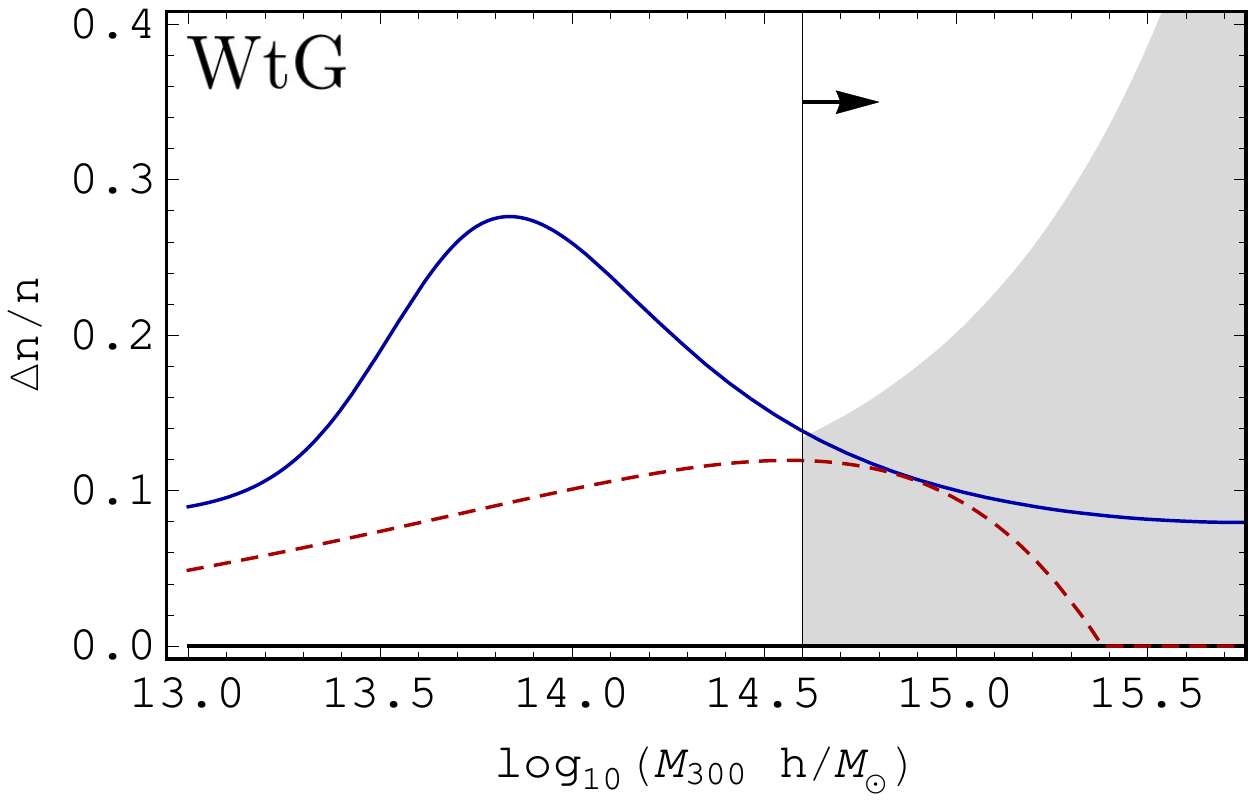}
\quad
\includegraphics[width=0.48\columnwidth]{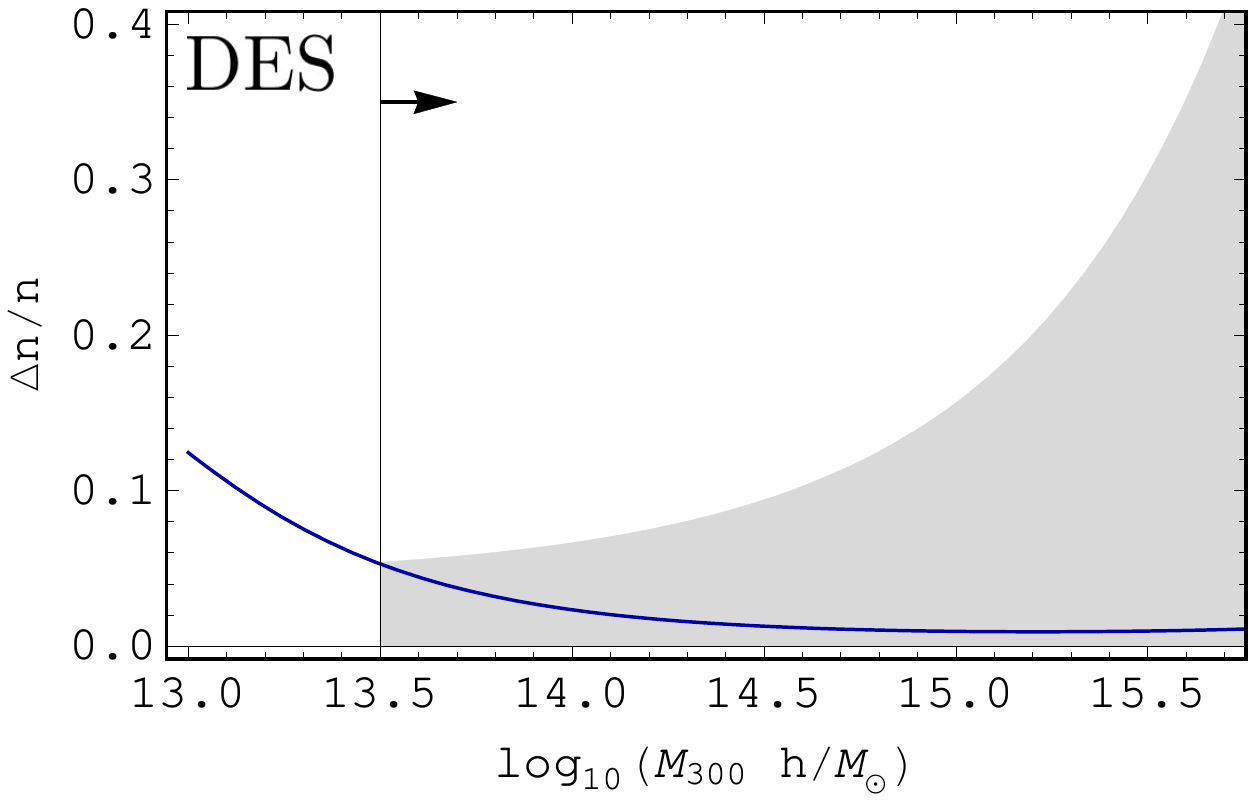}
\end{center}
\caption{Approximate forecast constraints on the enhancement of the growth of structure due to $f(R)$ gravity from available and forthcoming cluster abundance data. The blue lines are based on our new mass function modeling (see Eqs.~\eqref{eq:dceff}-\eqref{eq:correction}) and the grey shaded regions on the weak lensing mass calibration uncertainty $\epsilon_{\rm cal}$ for the fiducial $\Lambda$CDM cosmology of Eq.~\eqref{eq:fiducial_cosmo}. Vertical lines indicate the lowest cluster mass ends of each data set. \textit{Left:} the dashed red line shows the conservative mass function used in Cataneo et al.~(2015)~\cite{Cataneo:2014kaa} at $z=0$ for $|f_{R0}| = 1.62 \times 10^{-5}$, which corresponds to the 95.4\% upper limit constraint obtained there. The solid blue line was calculated using the new HMF of this work with $|f_{R0}| = 8 \times 10^{-6}$, which matches the mass calibration $\epsilon_{\rm cal} = 7\%$ from WtG (grey shaded area) and represents a factor of 2 improvement over the current result of the red line using the same data. \textit{Right:} using upcoming DES data down to lower mass objects with a low redshift limit of $z=0.1$ and $\epsilon_{\rm cal} = 5\%$ (grey shaded area), our new model promises an improvement of the upper bound on $f(R)$ gravity at cluster scales of an order of magnitude, $|f_{R0}| = 10^{-6}$ (blue line).}
\label{fig:forecast}
\end{figure}

We can now use our fits to make approximate forecasts of the maximum background scalaron field amplitude allowed by cluster abundance data from existing and ongoing surveys. Our fiducial cosmology is $\Lambda$CDM in standard GR with the parameters set to the mean values obtained from the full statistical analysis carried out in~\cite{Cataneo:2014kaa} for the data combination dubbed there \textit{Clusters+Planck+WP+lensing+ACT+SPT+SNIa+BAO}, namely 
\begin{equation}
(\Omega_{\rm m}, \Omega_\Lambda, h, n_s, \sigma_8) = (0.3, 0.7, 0.683, 0.963, 0.82).
\label{eq:fiducial_cosmo}
\end{equation}
We opt for a Tinker et al.~(2010)~\cite{Tinker:2010my} fiducial mass function, which we calculate using {\sc HMFcalc}\footnote{\url{http://hmf.icrar.org}}~\cite{Murray:2013qza} with the model parameters fitted at $\Delta = 300$. Also, for the current purpose we assume that the only observational error is the uncertainty on the weak lensing mass calibration~\cite{Sealfon:2006hx,Johnston:2005uu,Sheldon:2007ty,Mandelbaum:2007dp,Leauthaud:2009mg,White:2010pk,Rozo:2010ew,Applegate:2012kr,vonderLinden:2014haa,Applegate:2015kua} since this is presently the dominant source of error in measurements of the cluster mass function. We then estimate the uncertainty $\epsilon_{\rm MF}$ on the observed cluster number counts by propagating the lensing calibration error $\epsilon_{\rm cal}$ as
\bq
\epsilon_{\rm MF} = \left| \frac{d\log_{10} n_{\ln M}^{\rm Tinker}}{d\log_{10} M} \right| \epsilon_{\rm cal}\,,
\label{eq:hmferror}
\eq
where $n_{\ln M}^{\rm Tinker}$ represents the fiducial mass function. First, based on results from the Weighing the Giants (WtG) project we consider a mass calibration error of 7\%~\cite{Applegate:2012kr}. This data was used in~\cite{Cataneo:2014kaa} together with a conservative mass function model~\cite{Schmidt:2008tn} to determine the upper bound $|f_{R0}| < 1.62 \times 10^{-5}$ at 95.4\% confidence level. The left panel of Fig.~\ref{fig:forecast} shows the corresponding mass function model at $z=0$ for the cosmological parameters given in Eq.~\eqref{eq:fiducial_cosmo} and this upper limit of $|f_{R0}|$ (dashed red line). The grey shaded area in this panel is the region allowed by the mass calibration uncertainty. Matching simply by visual inspection the expected likelihood of our new model (solid blue line) to that of the previous, conservative model  promises a potential improvement over our current constraints of a factor of $\sim 2$, i.e. $|f_{R0}| \lesssim 8 \times 10^{-6}$. Looking a bit further ahead, the ongoing Dark Energy Survey (DES)~\cite{Abbott:2005bi} should achieve a mass calibration precision of at least 5\% in the coming years (Joerg Dietrich, private communication; see also~\cite{Melchior:2016mp}) and be able to provide a sample with objects down to masses $\sim 10^{13.5} \, \Msunh$ and redshifts $z \sim 0.1$. The right panel of Fig.~\ref{fig:forecast} suggests that with this data we could potentially reduce the current upper limit from Cataneo et al.~(2015)~\cite{Cataneo:2014kaa} by an order of magnitude, reaching a background Compton wavelength $\lambda_{\rm C} \approx 2 \, \hMpc$ or equivalently $|f_{R0}| \approx 10^{-6}$. Remarkably, this forecast at cluster scales is competitive with current constraints from local gravity tests. Assuming a galactic Navarro-Frenk-White halo density profile~\cite{Navarro:1995iw} embedded in the cosmological background, these tests require an active chameleon screening inside the Galaxy from the center out to the location of our Solar System~\cite{Hu:2007nk,Lombriser:2013eza} to suppress unobserved modifications above $|f_{R0}| \sim 10^{-6}$.

\section{Conclusions} \label{sec:conclusions}

The abundance of galaxy clusters is sensitive to the growth of the large scale structure, and as such can effectively test departures from GR on cosmological scales. Upcoming and future cluster surveys will provide exquisite data, requiring accurate percent level theoretical predictions to realize the full potential of these measurements. In this work we have presented a novel semi-analytical approach that combines the advantages of the spherical collapse model of Lombriser et al.~(2013)~\cite{Lombriser:2013wta} with the information available in fully nonlinear cosmological simulations. Taking GR as a baseline theory of gravity, we have calibrated mass function ratios in the context of $f(R)$ gravity and obtained fitting functions for our additional parameters able to predict these ratios within a 5\% precision for the ranges $10^{13.5} \leqslant M_{300 \rm m} (\Msunh)^{-1} \leqslant 10^{15.5} $, $10^{-6} \leqslant |f_{R0}| \leqslant 10^{-4}$ and $0 \leqslant z \leqslant 0.5$. This corresponds to about a $50\%$ improvement on the purely spherical collapse results of~\cite{Lombriser:2013wta}. A similar level of accuracy can be achieved for the full $f(R)$ mass function on the condition that the modeling of the reference GR halo abundance is accurate at the percent level. Although in Eqs.~\eqref{eq:HMFzfits}-\eqref{eq:thetafits} we provide fits only for halo masses defined by mean matter densities of $\bar{\rho}=300\rhomb$, our relations can be readily refitted using other mass definitions (e.g. $\bar{\rho}=500\bar{\rho}_{\rm cr}$) bearing in mind the resolution limitations imposed by Eq.~\eqref{eq:nhalomin}.

Our method can also be straightforwardly applied to calibrate theoretical mass functions of other scalar-tensor theories characterized by a mass and environment dependent spherical collapse threshold. This is for example the case of the dilaton and symmetron models investigated in Brax et al.~(2012)~\cite{Brax:2012sd}. Note also that baryonic physics is likely to currently be irrelevant for the HMF ratios~\cite{Cataneo:2014kaa,Puchwein:2013lza}, and that any departures from DM-only predictions due to baryons could be included through a baseline GR mass function calibrated against hydrodynamical simulations (see e.g.~\cite{Bocquet:2015pva}). Analogous considerations might hold as well for the impact of massive neutrinos on the $f(R)$ to GR halo mass function ratio. It would be interesting to test the performance of our method on cosmological simulations incorporating massive neutrinos in both GR and $f(R)$  (see e.g. Baldi et al.~(2014)~\cite{Baldi:2013iza}). If the accuracy of our predictions remains unchanged when allowing a varying effective sum of the neutrino masses, then it would be sufficient to implement the prescription of Castorina et al.~(2014)~\cite{Castorina:2013wga} on the baseline GR mass function. Finally, in addition to Poisson noise it will be necessary to account for the uncertainty due to sample variance in order to unbiasedly constrain the low mass end of the HMF with forthcoming cluster number count data~\cite{Hu:2002we}. For the specific cosmological models of interest, this will require the calculation of the linear bias parameter, which in itself depends on the spherical collapse threshold~\cite{Sheth:1999mn,Tinker:2010my}. For $f(R)$ gravity, we should be able to use our effective linearly extrapolated overdensity (see Eq.~\eqref{eq:dceff}) to evaluate the linear bias and hence the sample variance contribution (Cataneo et al., in preparation) needed for a series of upcoming key cosmological analyses from ongoing and future cluster surveys.


\section*{Acknowledgements}

This work used the DiRAC Data Centric system at Durham University, operated by the Institute for Computational Cosmology on behalf of the STFC DiRAC HPC Facility \break (www.dirac.ac.uk). This equipment was funded by BIS National E-infrastructure capital grant ST/K00042X/1, STFC capital grants ST/H008519/1 and ST/K00087X/1, STFC DiRAC Operations grant ST/K003267/1 and Durham University. DiRAC is part of the National E-Infrastructure. Further numerical computations have been performed with Wolfram $Mathematica^{\rm \tiny \textregistered}~9$. For part of this work, the Dark Cosmology Centre was funded by the Danish National Research Foundation. DR is supported by a NASA Postdoctoral Program Senior Fellowship at the NASA Ames Research Center, administered by the Universities Space Research Association under contract with NASA. LL is supported by a SNSF Advanced Postdoc.Mobility Fellowship (No. 161058) and the STFC Consolidated Grant for Astronomy and Astrophysics at the University of Edinburgh. BL is supported by STFC Consolidated Grant No. ST/L00075X/1 and No. RF040335.
%


\vfill
\bibliographystyle{JHEP}
\bibliography{references}

\end{document}